%% file: main.tex
\documentclass[sigconf]{acmart}\AtBeginDocument{%
  }

\usepackage{float}
\usepackage{microtype}
\usepackage{graphicx} 
\usepackage{makecell}
\usepackage{amsmath}

\usepackage{multirow}
\usepackage{subfigure}

\usepackage{algorithm}
\usepackage{algorithmic}
\setcopyright{rightsretained}
\usepackage{flushend}

\copyrightyear{2023}
\acmYear{2023}
\acmPrice{15.00}
\acmDOI{10.1145/3534678.3539119}
\acmISBN{978-1-4503-9385-0/22/08}

\begin{document}

\title{Contrastive Multi-view Framework for Customer \\ Lifetime Value Prediction}


\author{Chuhan Wu$^1$, Jingjie Li$^1$, Qinglin Jia$^1$, Hong Zhu$^2$,  Yuan Fang$^2$, Ruiming Tang$^1$}
\renewcommand{\authors}{Chuhan Wu, Jingjie Li,  Qinglin Jia, Hong Zhu, Yuan Fang, Ruiming Tang}
\affiliation{%
  \institution{$^1$Noah’s Ark Lab, Huawei \quad \quad $^2$Consumer Cloud Service Interactive Media BU, Huawei}
  \country{}
  }
\email{{wuchuhan1, lijingjie1, jiaqinglin2, zhuhong8, frank.fy, tangruiming}@huawei.com}

\renewcommand{\shortauthors}{Chuhan Wu et al.}

\begin{abstract}
Accurate customer lifetime value (LTV) prediction can help service providers optimize their marketing policies in customer-centric applications.
However, the heavy sparsity of consumption events and the interference of data variance and noise obstruct LTV estimation. 
Many existing LTV prediction methods directly train a single-view LTV predictor on consumption samples, which may yield inaccurate and even biased knowledge extraction.
In this paper, we propose a contrastive multi-view framework for LTV prediction, which is a plug-and-play solution compatible with various backbone models.
It synthesizes multiple heterogeneous LTV regressors with complementary knowledge to improve model robustness and captures sample relatedness via contrastive learning to mitigate the dependency on data abundance.
Concretely, we use a decomposed scheme that converts the LTV prediction problem into a combination of estimating consumption probability and payment amount. 
To alleviate the impact of noisy data on model learning, we propose a multi-view framework that jointly optimizes multiple types of regressors with diverse characteristics and advantages to encode and fuse comprehensive knowledge.
To fully exploit the potential of limited training samples, we propose a hybrid contrastive learning method to help capture the relatedness between samples in both classification and regression tasks. 
We conduct extensive experiments on a real-world game LTV prediction dataset and the results validate the effectiveness of our method. 
We have deployed our solution online in Huawei's mobile game center and achieved 32.26\% total payment amount gains.

\end{abstract}

%
%



\maketitle

\input{data/introduction.tex}

\input{data/relatedwork.tex}

\input{data/method.tex}

\input{data/experiment.tex}
\input{data/conclusion.tex}

\bibliographystyle{ACM-Reference-Format}
\bibliography{main}

\end{document}

%% file: data/introduction.tex
\section{Introduction}
Customer lifetime value (LTV) reflects the expected revenue contributed by a customer in a long-term relationship with a business~\cite{jain2002customer,gupta2006modeling}.
With the guidance of LTV in business planning and user segmentation, service providers can make more informed marketing decisions and provide better personalized services to increase user retention and reduce churn~\cite{berger1998customer,mani1999statistics,rosset2003customer,venkatesan2004customer,glady2009modeling,pollak2021predicting}. 
Thus, accurate LTV prediction is playing an increasingly important role in the digital economy and intelligent management~\cite{malthouse2005can,chamberlain2017customer,dahana2019linking,win2020predicting}.

\begin{figure}[!t]
  \centering
    \includegraphics[width=0.95\linewidth]{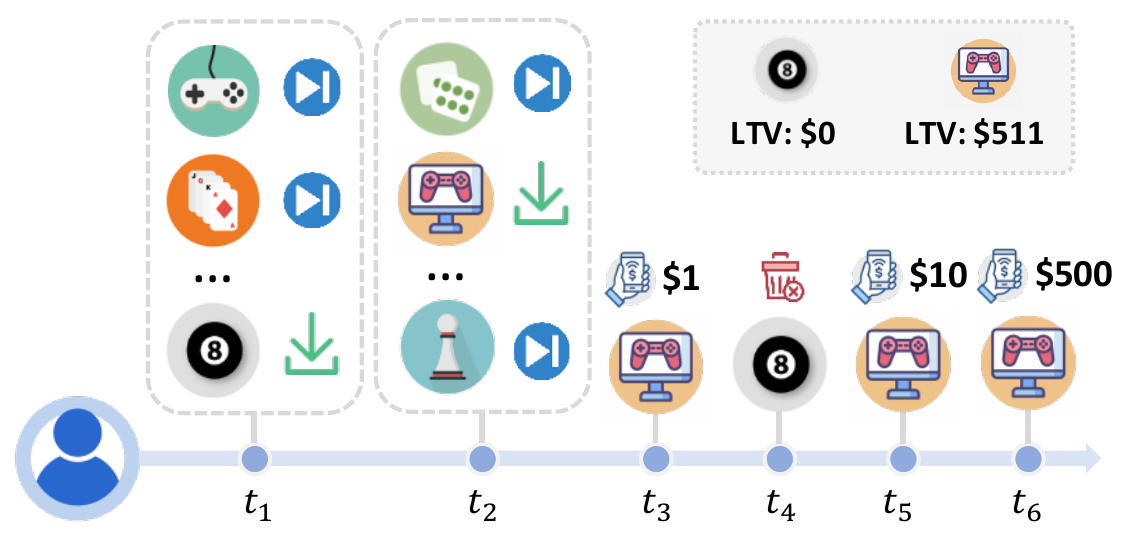}
  \caption{The consumption behavior patterns of a user in a typical App usage scenario.}
  \label{fig.exp}
\end{figure}

In many real-world applications, users' consumption behaviors are sparse, volatile, and noisy in their nature~\cite{xing2021learning,li2022billion}.
For example, as shown in Fig.~\ref{fig.exp}, in the App usage scenario, users may only choose a few Apps from the various candidates recommended by the App store, and they may order paid services in fewer Apps after trials.
In our practice, only about 2\% of App downloads are finally converted into valid payments within a month.
Furthermore, users' purchase behaviors may be affected by various subject and objective factors, such as unplanned purchase intent and sales promotion~\cite{blattberg2009customer,akram2018website,zhang2020research}.
As a result, we can observe a quite wide-range, long-tailed, and noisy LTV distribution (see our dataset statistics for an intuitive example).
The heavy sparsity and noise of user consumption samples make LTV prediction a rather challenging problem.

Considering the sparsity of samples with non-zero LTVs, many methods decompose the LTV prediction problem into a binary classification task to predict premium users and a regression task to estimate potential monetary values~\cite{wang2019deep}.
Some of them use two-stage strategies to model the two tasks in a cascaded way and use machine learning models such as random forests~\cite{vanderveld2016an} and XGBoost~\cite{drachen2018to} to process input features.
However, maintaining multiple models in these two-stage methods may yield higher complexity and heavier error propagation problem.
In recent years, several deep learning-based methods provide end-to-end solutions to LTV prediction.
For example, ZILN~\cite{wang2019deep} unifies the two tasks in a single model via multi-task learning, where the final predicted LTV is given by the multiplication between the purchase propensity and the expected purchase amount.
ODMN~\cite{li2022billion} models the temporal dependencies between LTVs, and uses a multi-distribution multi-expert module that divides the imbalanced LTV distribution into several smoother sub-distributions and assigns them to different experts.
However, the single-view LTV regressors in these methods may still be severely perturbed and even biased by the outlier and noisy samples, especially when there are only very limited positive samples for model optimization.

In this paper, we propose a \underline{C}ontrastive \underline{M}ulti-view framework for \underline{LTV} prediction named \textit{CMLTV}, which can effectively mitigate the impact of the consumption data sparsity, volatility, and noise on model learning.
The core of \textit{CMLTV} is a multi-view regression framework, where multiple heterogeneous regressors with diverse characteristics and advantages are jointly optimized to extract complementary and robust knowledge.
Their predictions are synthesized into a unified score, which is further combined with the purchase probability estimated by a purchase classifier to generate the final LTV.
To reduce the model's dependency on data volume, we propose a hybrid contrastive learning method to help capture the relatedness between samples in both classification and regression tasks to fully exploit data potentials.
By jointly regularizing the purchase classifier and multi-view regressors, the model can be aware of the inherent relatedness among samples and make more accurate predictions. 
Our offline experiments are conducted on a real-world game App user LTV prediction dataset collected from Huawei's mobile game center, which shows the effectiveness of \textit{CMLTV}. 
Further online A/B test on this platform also shows 33.26\% total payment amount gains.
Up to now, \textit{CMLTV} has been deeply involved to provide stable and high-quality game App suggestion services for hundreds of millions of users.

The contributions of this paper are listed as follows:
\begin{itemize}
    \item We propose a multi-view LTV prediction framework that can encode complementary knowledge via multiple heterogeneous regressors to confront data noise and imbalance. 
    \item We propose a hybrid contrastive learning method that enables the model to organically exploit the relatedness between samples in both classification and regression tasks to mitigate the impact of data sparsity.
    \item We conduct extensive experiments in both offline and online environments, and the results fully validate the effectiveness of our method.
    Our solution has been deployed online for personalized game App suggestions, serving hundreds of millions of mobile users.
\end{itemize}

%% file: data/relatedwork.tex
\section{Related Work}\label{sec:RelatedWork}

\subsection{LTV Prediction}

LTV prediction is a classic problem in business management~\cite{kumar2004customer}.
Pioneer research on customer LTV prediction focuses on building probabilistic models based on observed data~\cite{jasek2019comparative}.
For example, Pareto/NBD~\cite{schmittlein1987counting} is a canonical model for customer analysis based on historical transactions, which forecasts the future activity and purchase frequency of users by modeling user behaviors with certain stochastic processes.
Based on this framework, \citet{fader2005rfm} studied grouping users based on the recency, frequency, and monetary value of their transactions, and then estimated the LTV for each customer cohort.
These probability-based methods require prior knowledge of user behaviors to build basic model assumptions, which are highly sensitive to data noise and volatility.
In complex scenarios, it is also difficult to simulate user behaviors via well-established stochastic processes.

Another major research line of LTV prediction is using machine learning techniques to build data-driven models from historical logs~\cite{tkachenko2015autonomous,chen2018customer,sabbeh2018machine,del2019profiling,burelli2019predicting,jiang2020study,tekin2022customer}.
Considering the characteristics of LTV distribution, several methods decompose the task into two stages, i.e., stage one predicts the purchase propensity and stage two estimates the LTVs of users who are predicted to purchase in stage one.
For example, \citet{vanderveld2016an} proposed to use random forests in both stages, which are built on various features including user engagement measurements, user experience, historical behaviors, and demographics.
\citet{drachen2018to} proposed to use XGBoost models to implement a game LTV prediction framework that involves various gameplay, social, and purchase features.
These two-stage methods need to maintain multiple cascaded models, which may lead to high computational costs and online latency.

In recent years, several deep learning-based methods unify the two stages into a single model.
For example, ZILN~\cite{wang2019deep} uses a multi-task learning framework that optimizes a binary classifier to predict purchase propensity and a lognormal distribution-based regressor to predict the expected payment amount.
The final LTV is given by multiplying the predicted purchase propensity and payment amount.
ODMN~\cite{li2022billion} further considers the sequential dependencies between LTV in different periods.
At each prediction point, it uses a multi-distribution multi-expert module that predicts the classification probabilities of different LTV ranges and uses them to adaptively select proper experts to predict LTV in certain ranges.
Unfortunately, due to the noisy, imbalanced, and volatile nature of purchase data, the single-view LTV regressors in these methods may still be heavily affected and even biased, especially when there are insufficient samples to draw a panorama of data distribution.
Our work incorporates a multi-view regression framework to extract complementary information and fuses it into more robust and accurate knowledge.
It is remarkable that several works in recent years studied LTV prediction in other aspects, such as user representation~\cite{xing2021learning}, purchase sequence modeling~\cite{chen2018customer,bauer2021improved}, and social information mining~\cite{zhao2023percltv}.
Since these scenario-specific methods bring some additional constraints to data formats, they do not fall into the same research line as our work.

\begin{figure*}[!t]
  \centering
    \includegraphics[width=0.999\linewidth]{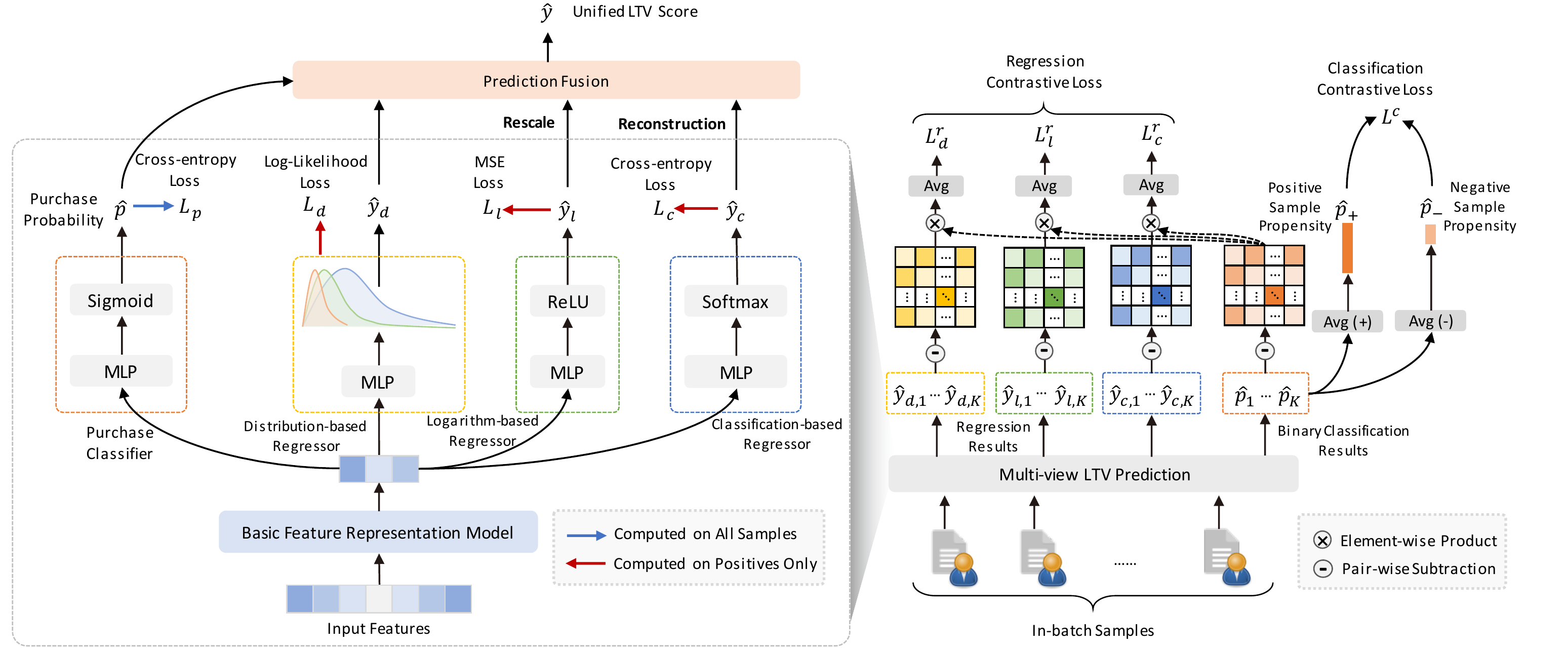}
  \caption{The framework of our \textit{CMLTV} method.}
  \label{fig.model}
\end{figure*}

\subsection{Contrastive Learning in Recommendation}

Contrastive learning is a widely employed technique in the recommendation field~\cite{wei2021contrastive,yu2022graph}, and here we briefly introduce a few relevant works.
A common contrastive paradigm in recommendation is comparing each positive sample with one or more negative samples, e.g., BPR~\cite{rendle2009bpr} and InfoNCE~\cite{oord2018representation} losses.
This paradigm is a popular choice in learning-to-rank systems in various recommendation scenarios, such as e-commerce~\cite{xie2022contrastive}, video~\cite{yi2022multi}, and news feed~\cite{wu2019npa}.
However, in online advertising applications, both the orders of candidate ads and the absolute values of predicted factors (e.g., pCVR and pLTV) matter in the bidding process~\cite{guo2017deepfm,pan2021click}.
Thus, the pair-wise contrastive learning paradigm may be inappropriate in advertising scenarios due to its distortion effects on predicted scores.
To adopt the strong representation ability of contrastive learning in advertising systems, \citet{wang2022cl4ctr} proposed a CL4CTR framework that uses data augmentation to construct contrast pairs.
However, it cannot effectively exploit the relatedness between different samples.
Moreover, most studies of contrastive learning in recommendation focus on classification problems~\cite{lin2022personalized,bai2022contrastive}, which are not fully compatible with LTV prediction that involves a regression task.
We propose a hybrid contrastive learning method that simultaneously guides model optimization in both classification and regression task, which is especially suitable for LTV prediction to alleviate the impact of data scarcity.

%% file: data/method.tex
\section{Methodology}\label{sec:Model}

Here we introduce our \textit{CMLTV} approach for customer LTV prediction.
Since there are several different formulations of LTV prediction in existing literature, we first present the definition of the problem studied in our paper.
We then describe our model in detail and show how to optimize it to generate LTV predictions.

\subsection{Problem Definition}

Denote a user and a certain product/service by $u$ and $i$, respectively.
Given the features of $u$ and $i$ (as well as some context features if available), the goal of an LTV prediction model is to estimate the payment amount of $u$ on $i$ within a certain time span $T$ (e.g., 30 days).
If the user $u$ does not purchase $i$ within $T$, the corresponding LTV in this period is labeled as zero. 
If there are multiple transactions concerning $u$ and $i$ in this time span, the LTV label is the summation of their monetary values.
After model learning on historical transaction data, it is expected to predict future LTVs as a reference for decision making and personalized advertising.

\subsection{CMLTV Framework}

The overall framework of our \textit{CMLTV} method is shown in Fig.~\ref{fig.model}.
When a batch of training samples arrives, we first use a multi-view LTV prediction model to generate the purchase probability and multiple LTV regression results for each sample, and then apply a hybrid contrastive learning strategy to samples in this batch to capture their inherent relatedness.
Their details are described below.

\subsubsection{Multi-View LTV Prediction}

The multi-view LTV prediction module uses heterogeneous regressors\footnote{The MLP networks in regressors may contain one or more layers, but we denote them as ``MLP'' for consistency.} with diverse characteristics to profile the input sample in different aspects.
Denote the input features of a sample by $\mathbf{x}$.\footnote{We assume that discrete features have been converted into embeddings.}
Firstly, a basic feature representation model is used to learn the hidden representations $\mathbf{h}$ of input features by modeling their interactions.
Note that our framework does not restrict the architecture of this backbone model, and it can be implemented by various off-the-shelf structures such as MLP, DeepFM~\cite{guo2017deepfm}, DCN~\cite{wang2017deep}, and DCNv2~\cite{wang2021dcn}.

Next, a purchase classifier is applied to the hidden representation $\mathbf{h}$ to predict the purchase probability $\hat{p}$ as follows:
\begin{equation}
\begin{aligned}
\mathbf{h}_p &= \rm{ReLU}(\mathbf{W}_p\mathbf{h}+\mathbf{b}_p), \\
    \hat{p} &= \sigma(\mathbf{w}_p\mathbf{h}_p+b_p),
    \end{aligned}
\end{equation}
where $\mathbf{W}_p$, $\mathbf{w}_p$, $\mathbf{b}_p$, and $b_p$ are parameters, and $\sigma(\cdot)$ stands for the sigmoid function.
The probability output by the classifier is used to compute the cross-entropy loss $\mathcal{L}_p$ as follows:
\begin{equation}
    \mathcal{L}_p = -z\log\hat{p}-(1-z)\log{(1-\hat{p})},
\end{equation}
where $z$ is the binary purchase label of a sample.

We then estimate the payment amount of each sample in a multi-view manner.
Here we include three different types of regressors.
The first one is a distribution-based regressor, which aims to approximate real LTV distributions with well-formulated probabilistic models.
Although it can be implemented by any probability distributions, we recommend using some canonical ones such as lognormal and gamma distributions that have been shown effective in characterizing long-tail property~\cite{kim2008selection,glady2009modified,yang2022personalized}.
Here we denote the normalized probability density function of the selected distribution by $f(x)$.
It is parameterized by a variable\footnote{It can be either a scalar or a vector, depending on the parameters required by $f(x)$.} $\mathbf{\theta}$, which is learned from $\mathbf{h}$ as follows:
\begin{equation}
\begin{aligned}
\mathbf{h}_d &= {\rm{ReLU}}(\mathbf{W}_d\mathbf{h}+\mathbf{b}_d), \\
    \mathbf{\theta}' &= \mathbf{w}_d\mathbf{h}_d+b_d,\\
   \mathbf{\theta} &=  \log[1+\exp(\mathbf{\theta}')],
    \end{aligned}
\end{equation}
where $\mathbf{W}_d$, $\mathbf{w}_d$, $\mathbf{b}_d$, and $b_d$ are learnable parameters.
The loss function $\mathcal{L}_d$ of this regressor is the negative log-likelihood of the probability density given by $f(x)$, which is formulated as follows:
\begin{equation}
    \mathcal{L}_d = - \log f(y),
\end{equation}
where $y$ is the LTV label of a sample.
By optimizing this loss function, the model is encouraged to find better   $\mathbf{\theta}$ to parameterize $f(x)$ that can maximize the LTV probability density of training samples.
In the test phase, it uses the expectation of the distribution $f(x)$ parameterized by the $\mathbf{\theta}$ obtained from each sample as the prediction.
In our approach, we use gamma distribution to instantiate the function $f(x)$.
Therefore, the variable $\mathbf{\theta}$ is a vector including two elements, i.e., a shape parameter and a rate parameter, and the predicted LTV score is the division between them.

The second regressor is logarithm-based, which aims to predict the LTV score in the log scale.
Although customer LTVs can range from zero to million-level numbers, they can be only low single digits in the logarithm view, which are suitable for common neural networks to handle.
Concretely, the output $\hat{y}_l$ of this regressor is obtained from $\mathbf{h}$ through an MLP module as follows: 
\begin{equation}
\begin{aligned}
\mathbf{h}_l &= {\rm{ReLU}}(\mathbf{W}_l\mathbf{h}+\mathbf{b}_l), \\
    \hat{y}_l &= {\rm{ReLU}}(\mathbf{w}_l\mathbf{h}_l+b_l),
    \end{aligned}
\end{equation}
where $\mathbf{W}_l$, $\mathbf{w}_l$, $\mathbf{b}_l$, and $b_l$ are MLP parameters.
The loss function $\mathcal{L}_l$ of this regressor is the standard Mean Squared Error (MSE), which is computed as follows:
\begin{equation}
    \mathcal{L}_l = (\hat{y}_l - \lg(1+y))^2.
\end{equation}
Here we add one to the label to ensure the non-negativity.

The third regressor is classification-based, which first converts the regression task into a classification problem and then reconstructs a real-valued score from the class probabilities as the final prediction.
Since LTV distribution is usually long-tailed, it divides the whole range into several class bins using logarithm operations.
We use base 2 logarithms for binning rather than larger bases because it can keep more fine-grained numerical information.\footnote{While it is possible to choose smaller log bases to achieve finer-grained binning, it may lead to an excess of class numbers and even redundant classes.}
More specifically, the corresponding class ID $c$ of LTV label $y$ is obtained as follows:
\begin{equation}
    c = \lfloor \log_2 (1+y) \rfloor.
\end{equation}
The class probability vector $\mathbf{\hat{y}}_c$ is predicted as follows:
\begin{equation}
\begin{aligned}
\mathbf{h}_c &= {\rm{ReLU}}(\mathbf{W}_c\mathbf{h}+\mathbf{b}_c), \\
    \mathbf{\hat{y}}_c &= {\rm{Softmax}}(\mathbf{V}_c\mathbf{h}_c+\mathbf{v}_c),
    \end{aligned}
\end{equation}
where $\mathbf{W}_c$, $\mathbf{V}_c$, $\mathbf{b}_c$, and $\mathbf{v}_c$ are classifier parameters.
The loss function $\mathcal{L}_c$ for model learning is the following multi-class cross-entropy:
\begin{equation}
    \mathcal{L}_c =  - \sum_{i=1}^C c_i \log \mathbf{\hat{y}}_c^i,
\end{equation}
where $C$ is the class number, $c_i$ and $\hat{y}_c^i$ is the real and predicted label for the $i$-th class, respectively.
The final real-valued LTV prediction $\hat{y}'_c$ is reconstructed from the probability vector $\mathbf{\hat{y}}_c$ as follows:
\begin{equation}
    \hat{y}'_c = \sum_{i=1}^C \frac{(2^i-1+2^{i+1}-2)}{2} \mathbf{\hat{y}}_c^i = \sum_{i=1}^C \frac{(2^i + 2^{i+1}-3)}{2} \mathbf{\hat{y}}_c^i.
\end{equation}
In this way, we use the expectation as LTV prediction, where the center of each bin is used to represent its expected LTV.

\subsubsection{Hybrid Contrastive Learning}

Next, we introduce the hybrid contrastive learning mechanism in our approach, which aims to build organic connections between samples in the same batch to better exploit the potential of limited training data.
Assume there are $K$ samples within a batch.
Their predicted purchase probabilities and three types of LTV scores are denoted by $[\hat{p}_1, ..., \hat{p}_K]$, $[\hat{y}_{d,1}, ..., \hat{y}_{d,K}]$,  $[\hat{y}_{l,1}, ..., \hat{y}_{l,K}]$, $[\hat{y}_{c,1}, ..., \hat{y}_{c,K}]$, respectively.
Inspired by BPR loss, we encourage the purchase probabilities of positive samples to be larger than negative samples.
However, it is infeasible to directly compare each pair of samples because the noisy samples may mislead the model to generate low-quality contrastive pairs, which are harmful to model optimization.
Fortunately, although parts of the samples are uninformative, positive samples should be assigned higher purchase propensities than negative samples on average.
Thus, we first compute the average purchase probabilities of positive and negative samples, which are denoted by $\hat{p}_+$ and $\hat{p}_-$, respectively.
We design a classification contrastive loss $\mathcal{L}^c$ by comparing $\hat{p}_+$ and $\hat{p}_-$ as follows:
\begin{equation}
    \mathcal{L}^c = -\log\sigma[\sigma^{-1}(\hat{p}_+) - \sigma^{-1}(\hat{p}_-)],
\end{equation}
where $\sigma^{-1}$ means the inverse function of sigmoid.

Motivated by the findings in prior work~\cite{borle2008customer}, we assume that users with higher LTVs may also have higher purchase probabilities.
This assumption is probably true in various scenarios because large consumption values are usually caused by multiple purchase behaviors.
Thus, we propose a regression contrastive learning method to regularize the regression results to be positively correlated with the predicted purchase probabilities.
We take the scores output by the distribution-based regressor as an example to elaborate this process.
Its corresponding regression contrastive loss $\mathcal{L}^r_d$ is computed as follows:
\begin{equation}
    \mathcal{L}^r_d = - \frac{1}{K^2}\sum_{i=1}^K\sum_{j=1}^K (\hat{p}_i-\hat{p}_j)[\lg(1+\hat{y}_{d,i})-\lg(1+\hat{y}_{d,j})].
\end{equation}
Here we use base 10 logarithms to control the scale of contrastive loss.
Similarly, we compute the regression contrastive losses $\mathcal{L}^r_l$ and $\mathcal{L}^r_c$ based on the other two types of regression results.
In this way, the relatedness between different samples in the same batch is explicitly encoded into model learning, meanwhile the classification and regression parts are naturally connected by these losses so that their encoded knowledge can be exchanged and shared.

\subsection{Model Training and Prediction}

Finally, we introduce the details of model training and testing in our framework.
Our method uses a multi-task learning framework to unify different types of loss functions.
The overall loss $\mathcal{L}$ for model training is a combination of the binary classification loss, three types of regression losses, contrastive classification loss, and three types of contrastive regression losses.
Note that the binary classification loss and the contrastive losses are computed on all samples, while the regression losses are only activated on positive samples.
This is because the heavy data imbalance shall damage the accuracy of regression models.
Therefore, the loss $\mathcal{L}$ is formulated as follows:
\begin{equation}
\mathcal{L} = \mathcal{L}^c + \mathcal{L}^r_d + \mathcal{L}^r_l + \mathcal{L}^r_c + \sum_i [\mathcal{L}_p (i)] + \sum_{y_i>0} [\mathcal{L}_d (i)+\mathcal{L}_l (i)+\mathcal{L}_c (i)], 
\end{equation}
where $\mathcal{L}_p (i)$, $\mathcal{L}_d (i)$, $\mathcal{L}_l (i)$, and $\mathcal{L}_c (i)$ represent the corresponding loss on the $i$-th sample, and $y_i$ is its LTV label.
By optimizing this objective function, the model can fuse the knowledge extracted by the multi-view framework to generate accurate predictions.

After model convergence, we use the multi-view LTV prediction part for inference.
Motivated by the ZILN~\cite{wang2019deep} framework, we multiply the purchase probability and the regression scores.
The final LTV score $\hat{y}$ is generated as follows:
\begin{equation}
    \hat{y} = \hat{p}\cdot  [\alpha \hat{y}_d + \beta \hat{y}_l + (1-\alpha-\beta) \hat{y}_c],
\end{equation}
where $\alpha$ and $\beta$ are weighting coefficients.\footnote{We do not fuse the three scores in the model training phase to learn these coefficients because it will lead to performance degradation.}

\subsection{Model Complexity Analysis}

Finally, we provide some analysis of the theoretical computational complexity of our approach.
The complexity of the multi-view LTV prediction part mainly depends on the architecture of the basic model.
If it is implemented by an MLP model, the computational cost of this module is $O(Kd^2)$, where $d$ is the hidden dimension.
The computational cost of our hybrid contrastive learning part is mainly brought by computing the regression contrastive loss, whose complexity is $O(K^2)$.
Thus, the overall computational complexity of our framework is $O(Kd^2 + K^2)$.
If the batch size $K$ is much larger than the hidden dimension $d$, then the bottleneck is the contrastive learning part.\footnote{In our experiments, the contrastive learning mechanism actually occupies around 58\% of the training time.}
Therefore, it is preferable to set a moderate batch size due to efficiency considerations.

%% file: data/experiment.tex
\section{Experiments}\label{sec:Experiments}

\begin{figure}[!t]
	\centering
		\includegraphics[width=0.75\linewidth]{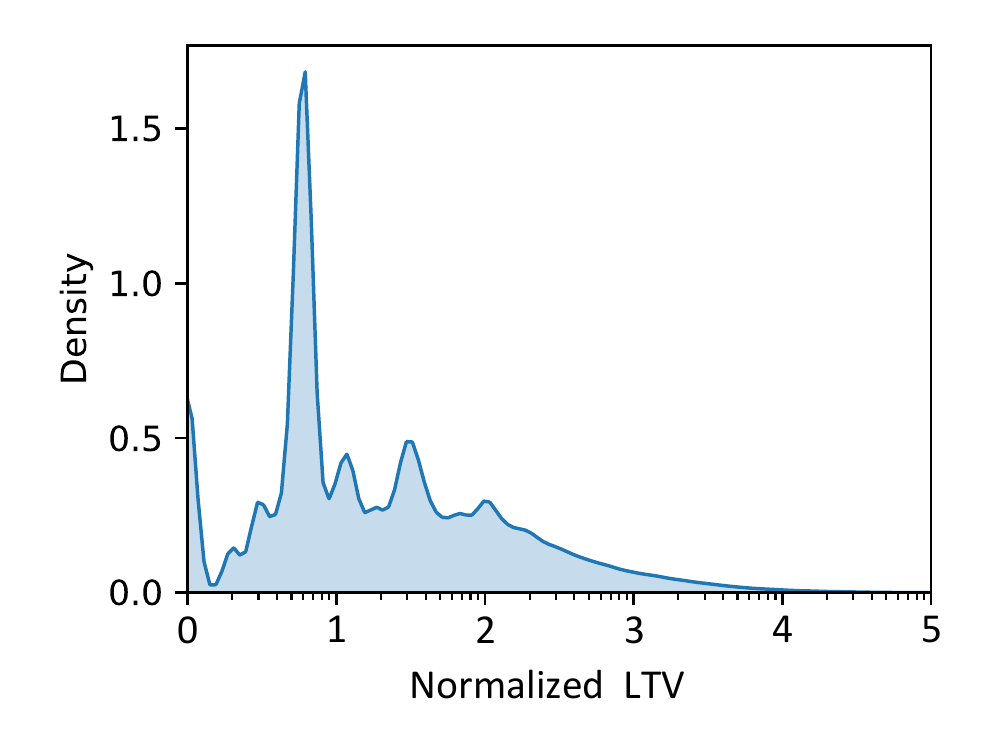}

\caption{The distribution of non-zero LTVs in our dataset.}\label{fig.ex1}
\end{figure}

\subsection{Dataset}

Although there are a few public datasets for LTV prediction, they may not well reflect the characteristics of many scenarios where positive samples are much sparser than negatives.\footnote{Some e-commerce datasets only contain users' transaction histories (i.e., positive samples), where negative samples are not included.}
Thus, we collect a dataset from Huawei's mobile game center for offline experiments.
It is sampled from the aggregated App consumption logs from Aug. 1st to Aug. 31st, 2022, and it contains 642k positive samples and 30.1m negative samples in total.
The LTV labels are counted within the 30 days after download.
For example, the LTV label of a sample generated on Aug. 1st is the accumulated consumption amount from Jul. 2nd to Aug. 1st.
Each sample is associated with several numerical features (e.g. historical LTVs), categorical features (e.g., App IDs and categories), and binary features (e.g., user segments). 
The distribution of non-zero LTVs (the values are normalized in the log scale) is illustrated in Fig.~\ref{fig.ex1}.
We find it approximately obeys a long-tail distribution but there are multiple peaks on the curve, which is possibly caused by the institutionalized price levels on mobile App stores. 
The samples on the last day are used for test and we randomly sample 10\% of training data for validation.

\begin{table*}[t]
\centering
\caption{LTV prediction performance of different methods on all samples or positive samples only.}\label{table.perf} 
\resizebox{1.0\linewidth}{!}{

\begin{tabular}{l|cccccc|ccccc}
\Xhline{1.0pt}
\multicolumn{1}{c|}{\multirow{2}{*}{\textbf{Methods}}} & \multicolumn{6}{c|}{\textbf{All Samples}}                                                & \multicolumn{5}{c}{\textbf{Positive Samples}}                                   \\ \cline{2-12} 
\multicolumn{1}{c|}{}                                  & RMSE            & MAE   & Pearson & Spearman        & R2\_score        & AUC             & RMSE            & MAE    & Pearson         & Spearman        & R2\_score        \\ \hline
Linear                                                 & 247.16          & 9.809 & 0.1422  & 0.1194          & 0.01958          & 0.7517          & 1770.4          & 236.85 & 0.3229          & 0.2927          & 0.02588          \\
MLP                                                    & 248.11          & 20.42 & 0.1257  & 0.0740          & 0.01209          & 0.6549          & 1789.0          & \textbf{232.96} & 0.3249          & 0.4772          & 0.00538          \\
RF                                         & 248.87          & 4.774 & 0.0949  & 0.1352          & 0.00599          & 0.7314          & 1803.0          & 241.59 & 0.2492          & 0.0949          & 0.01035          \\
XGBoost                                                & 246.40          & \textbf{4.681} & \textbf{0.2548}  & 0.1581          & 0.02565          & 0.8336          & 1786.3          & 241.43 & 0.2664          & 0.2424          & 0.00835          \\
ZILN                                                   & 242.56          & 8.090 & 0.2421  & 0.1708          & 0.05576          & 0.8601          & 1713.0          & 241.37 & 0.3456          & 0.4345          & 0.07721          \\
MDME                                                   & 243.71          & 7.969 & 0.2346  & 0.1668          & 0.04395          & 0.8482          & 1743.6          & 243.59 & 0.3286          & 0.4299          & 0.07843          \\ \hline
CMLTV                                                  & \textbf{240.16} & 6.889 & 0.2495  & \textbf{0.1724} & \textbf{0.07436} & \textbf{0.8629} & \textbf{1710.5} & 235.74 & \textbf{0.3746} & \textbf{0.4382} & \textbf{0.10291} \\ \Xhline{1.0pt}

\end{tabular}
}
\end{table*}

\begin{table*}[t]
\caption{Comparison of different frameworks in empowering different models.}\label{table.perf2} 
\resizebox{1.0\linewidth}{!}{

\begin{tabular}{l|cccccc|ccccc}
\Xhline{1.0pt}
\multicolumn{1}{c|}{\multirow{2}{*}{\textbf{Methods}}} & \multicolumn{6}{c|}{\textbf{All Samples}}                         & \multicolumn{5}{c}{\textbf{Positive Samples}}             \\ \cline{2-12} 
\multicolumn{1}{c|}{}                         & RMSE   & MAE   & Pearson & Spearman & R2\_score & AUC    & RMSE   & MAE    & Pearson & Spearman & R2\_score \\ \hline
MLP+ZILN                                      & 242.56 & 8.090 & 0.2421  & 0.1708   & 0.05576   & 0.8601 & 1713.0 & 241.37 & 0.3456  & 0.4345   & 0.07721   \\
MLP+MDME                                      & 243.71 & 7.969 & 0.2346  & 0.1668   & 0.04395   & 0.8482 & 1743.6 & 243.59 & 0.3286  & 0.4299   & 0.07843   \\
MLP+CMLTV                                     & 240.16 & 6.889 & 0.2495  & 0.1724   & 0.07436   & 0.8629 & 1710.5 & 235.74 & 0.3746  & 0.4382   & 0.10291   \\ \hline
DCN+ZILN                                      & 242.14 & 7.932 & 0.2425  & 0.1713   & 0.05685   & 0.8614 & 1704.2 & 239.89 & 0.3467  & 0.4348   & 0.07895   \\
DCN+MDME                                      & 243.53 & 7.885 & 0.2367  & 0.1689   & 0.04647   & 0.8525 & 1739.8 & 242.13 & 0.3305  & 0.4304   & 0.07822   \\
DCN+CMLTV                                     & 239.95 & 6.842 & 0.2510  & 0.1740   & 0.07569   & 0.8643 & 1699.6 & 234.36 & 0.3757  & 0.4390   & 0.10441   \\ \hline
AutoInt+ZILN                                  & 242.34 & 7.994 & 0.2420  & 0.1711   & 0.05664   & 0.8615 & 1706.7 & 241.01 & 0.3449  & 0.4340   & 0.07599   \\
AutoInt+MDME                                  & 243.69 & 7.910 & 0.2342  & 0.1672   & 0.04592   & 0.8509 & 1741.4 & 242.96 & 0.3278  & 0.4284   & 0.07644   \\
AutoInt+CMLTV                                 & 240.02 & 6.856 & 0.2494  & 0.1730   & 0.07498   & 0.8633 & 1702.3 & 234.98 & 0.3751  & 0.4377   & 0.10284   \\ \Xhline{1.0pt}
\end{tabular}
}
\end{table*}

\subsection{Experimental Settings}

In our experiments, we use a two-layer MLP with ReLU activation functions as the basic feature representation model in all deep learning methods, if not specified.
Each layer in the backbone model is followed by a batch normalization~\cite{ioffe2015batch} operation.
The hidden units of this model are 256 and 128, respectively.
The hidden dimension in the classifier and regressors is 64.
We use Adam~\cite{kingma2014adam} as the model optimizer and its learning rate is 1e-3.
The batch size is set to 10,000.
We use the early stopping mechanism in model training and its patient is 3 epochs.
The coefficients $\alpha$ and $\beta$ are set to 0.3.
These hyperparameters are tuned on the validation set.

Although many previous works~\cite{wang2019deep,li2022billion} use Rooted Mean Squared Error (RMSE) and Mean Absolute Error (MAE)\footnote{We do not use normalized RMSE and MAE because we find the data is so imbalanced that even normalized RMSE and MAE are larger than 1. Thus, we directly use the original non-normalized values. } as the performance metric, in our practice we find they are insufficient to comprehensively reflect prediction quality, especially when negative samples are dominant.
Thus, in our experiments we use six metrics to comprehensively evaluate model performance, including RMSE, MAE, Pearson correlation coefficient (denoted by Pearson), Spearman correlation coefficient (denoted by Spearman), coefficient of determination  (denoted by R2\_score), and Area Under ROC Curve (AUC).\footnote{The AUC score is evaluated by the binary purchase label.}
To fully understand model performance on different types of samples, we compute all metrics except AUC on all samples or positive samples only, respectively.
This is because the errors on positive samples and negative samples may yield  different volumes of loss in practical scenarios.
We repeat each experiment five times to mitigate randomness and report the average performance.

\subsection{Performance Evaluation}

In our offline evaluation, we compare our \textit{CMLTV} approach with the following baselines:
\begin{itemize}
    \item \textit{Linear}, which uses a multivariate linear regression model for LTV prediction;
    \item \textit{MLP}, which uses a multi-layer perceptron network for LTV prediction;
    \item \textit{RF}~\cite{vanderveld2016an}, which uses two independent random forest models for classification and regression;
    \item \textit{XGBoost}~\cite{drachen2018to}, replacing the random forest with the XGboost model in the above two-stage method;
    \item \textit{ZILN}~\cite{wang2019deep}, a multi-task learning method that unified binary classification and lognormal distribution-based regression;
    \item \textit{MDME}~\cite{li2022billion}, a multi-distribution multi-expert method for LTV prediction, which is the core module in ODMN~\cite{li2022billion}.
\end{itemize}
The performance of compared methods is listed in Table~\ref{table.perf}, from which we have the following observations:

\begin{figure*}[!t]
	\centering
	\includegraphics[width=0.99\linewidth]{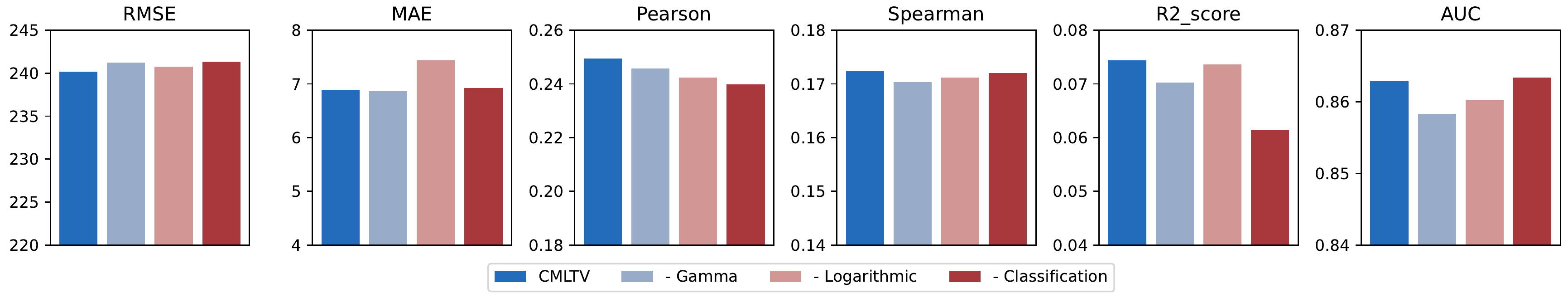}
	
\caption{The performance of \textit{CMLTV} and its variants without different regression views.}\label{fig.ex2}
\end{figure*}

\begin{figure*}[!t]
	\centering
	\includegraphics[width=0.99\linewidth]{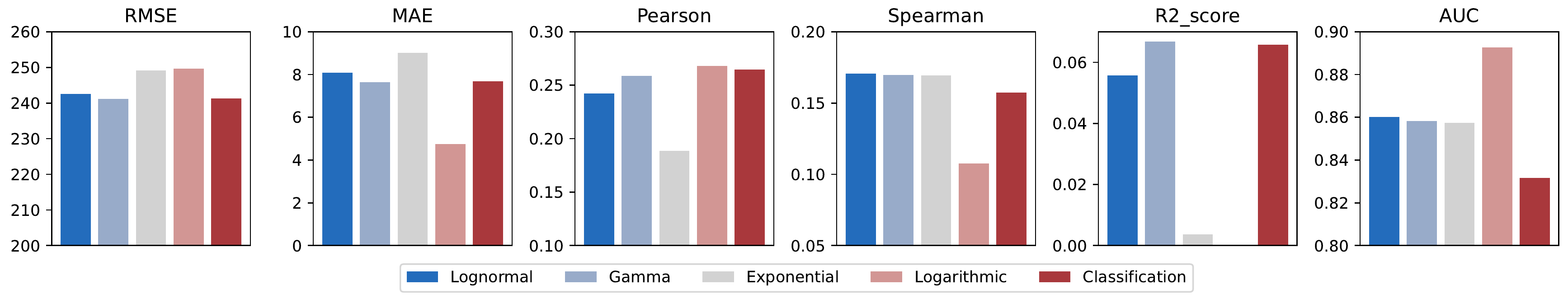}
	
\caption{The performance of using different types of regressors alone.}\label{fig.ex3}
\end{figure*}

First, we find different methods may have different ranks in terms of different metrics.
For example, on all samples RF achieves a low MAE score but performs poorly in terms of AUC.
The mismatch between different metrics is mainly due to the high data imbalance and the perturbation of outlier data.
Thus, it is insufficient to merely use RMSE and MAE to measure model performance.
Only if a model performs better than other models in most metrics, we can be  confident about its effectiveness.
According to the long-term experience in our practical scenarios, R2\_score usually has the best correlation with real online performance. 
RMSE and MAE are less informative since they are highly sensitive to outliers. 
Industrial practitioners can choose it as the primary metric if using multiple metrics is not supported.
Second, the \textit{Linear} and \textit{MLP} models perform poorly in many metrics. 
This is because they are heavily perturbed and even biased by the extremely imbalanced data.
It seems that \textit{MLP} is more easily to be affected by data imbalance than the simple linear regression model, which may be because of the high non-linearity of deep networks.
This shows the rationality of decomposing the LTV prediction problem into a classification task and a regression task in deep learning-based methods.
Third, single-stage deep learning methods such as \textit{ZILN} and \textit{MDME} usually outperform two-stage methods (i.e., \textit{RF} and \textit{XGBoost}).
This is because the separate models in the two stages cannot share their mutual knowledge and the errors produced by the first model are completely inherited by the second one.
Thus, unifying the two processes in a single model is a better choice.
Forth, our \textit{CMLTV} method outperforms other baselines (at the significant level of $p<0.05$ in two-sided t-test, evaluated on the best-performed metrics).
This may be because our approach better defends data noise and volatility meanwhile being aware of samples' relationships to overcome data sparsity.
Thus, our method is more robust and accurate than baselines.

Since \textit{ZILN}, \textit{MDME}, and \textit{CMLTV} are independent on the base model, we use them to empower several widely used base models, including MLP, DCN~\cite{wang2017deep}, and AutoInt~\cite{song2019autoint}.
The results are shown in Table~\ref{table.perf2}.
We observe that \textit{CMLTV} consistently outperforms \textit{ZILN} and \textit{MDME}, and the DCN based model achieves the best results.
This further verifies the effectiveness and generality of our approach in leveraging existing models that are well crafted for other tasks such as CTR prediction in LTV prediction.

\subsection{Effectiveness of Regression Views}

In this section we verify the effectiveness of different regression views in our multi-view LTV prediction module.
We compare the results of \textit{CMLTV} with its ablations with different regression views removed, as shown in Fig.~\ref{fig.ex2}.
In our later experiments, we only report the results on all samples due to space limitations.
We find it interesting that different regressors have different advantages in  view of different metrics.
For example, the AUC substantially declines when the gamma distribution-based regressor is removed, while the Pearson correlation coefficient degrades the most when the classification-based regressor is removed.
This phenomenon confirms that different regression views indeed encode heterogeneous and complementary knowledge, and none of them is redundant.
Therefore, combining the predictions from multi-view regressors can keep satisfactory performance in all metrics.

We then report the results of using different types of regressors alone in our framework to support the choice of regressors in our approach. 
Here we compare five different regressors, including: (1) lognormal distribution-based regressor in ZILN; (2) gamma distribution-based regressor in our method; (3) negative exponential distribution-based regressor; (4) logarithm-based regressor in our method; (5) classification-based regressor in our method, and the results are shown in Fig.~\ref{fig.ex3}.
We find gamma distribution is more suitable than lognormal and exponential distributions for LTV modeling in our scenario, which is probably because the shape of distribution curve shown in Fig.~\ref{fig.ex1} better matches the shape of gamma distribution.
Thus, we select gamma distribution-based regressor in our method.
In addition, logarithm-based and classification-based regressors show advantages and disadvantages in different metrics, thereby they can be complementary when being optimized jointly.
Based on the above analysis, we combine these regressors in our multi-view framework to comprehensively empower our model.

\subsection{Effectiveness of Hybrid Contrastive Learning}

\begin{figure*}[!t]
	\centering
	\includegraphics[width=0.99\linewidth]{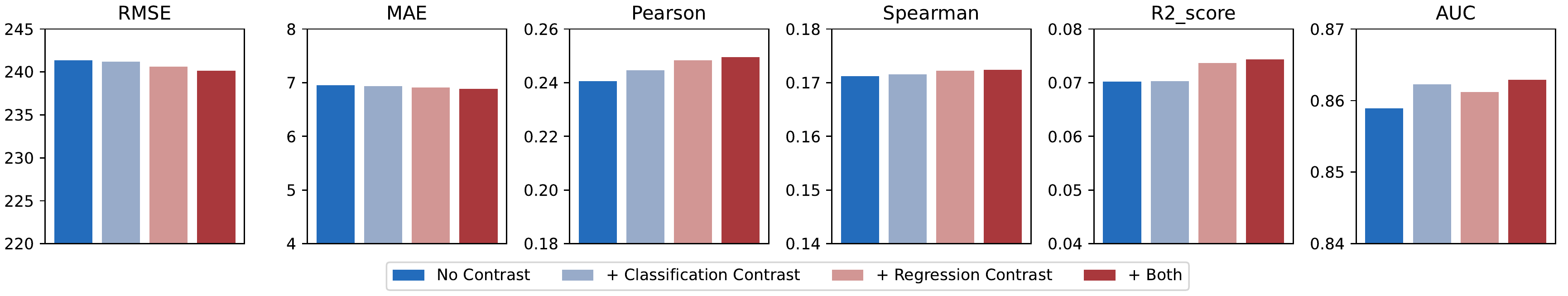}
	
\caption{Effectiveness of the two contrastive learning losses in \textit{CMLTV}.}\label{fig.ex4}
\end{figure*}

In this section we analyze the influence of the two types of contrastive learning losses in our approach.
The results of using different contrastive loss combinations in our method are shown in Fig.~\ref{fig.ex4}.
We observe that both losses are beneficial for model learning, while they show different effects on model performance.
The classification contrastive learning loss mainly improves AUC, which is intuitive because it is applied to the binary classification task.
On the contrary, the regression contrastive loss can substantially improve both AUC and other regression metrics.
This is because it builds connections between predicted purchase probabilities and LTV scores, thereby both classification and regression tasks can benefit from the mutual knowledge they encode.
Moreover, combining both tasks in our method can consistently achieve better results, which validates the effectiveness of our hybrid contrastive learning method.

\begin{figure*}[!t]
	\centering
	\includegraphics[width=0.99\linewidth]{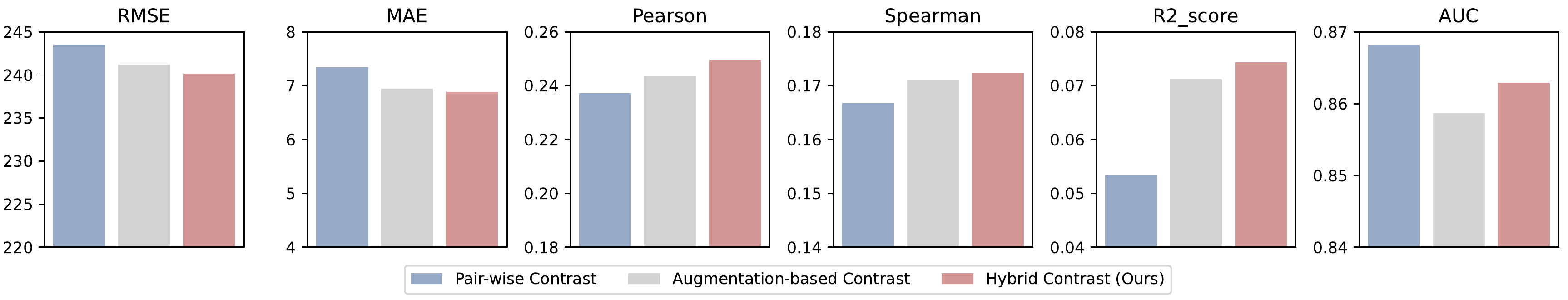}
	
\caption{Comparisons among different types of contrastive learning paradigms.}\label{fig.ex5}
\end{figure*}

\begin{figure}[!t]
	\centering
\subfigure[Purchase probabilities (y-axis is in log-scale).]{
	\includegraphics[width=0.47\linewidth]{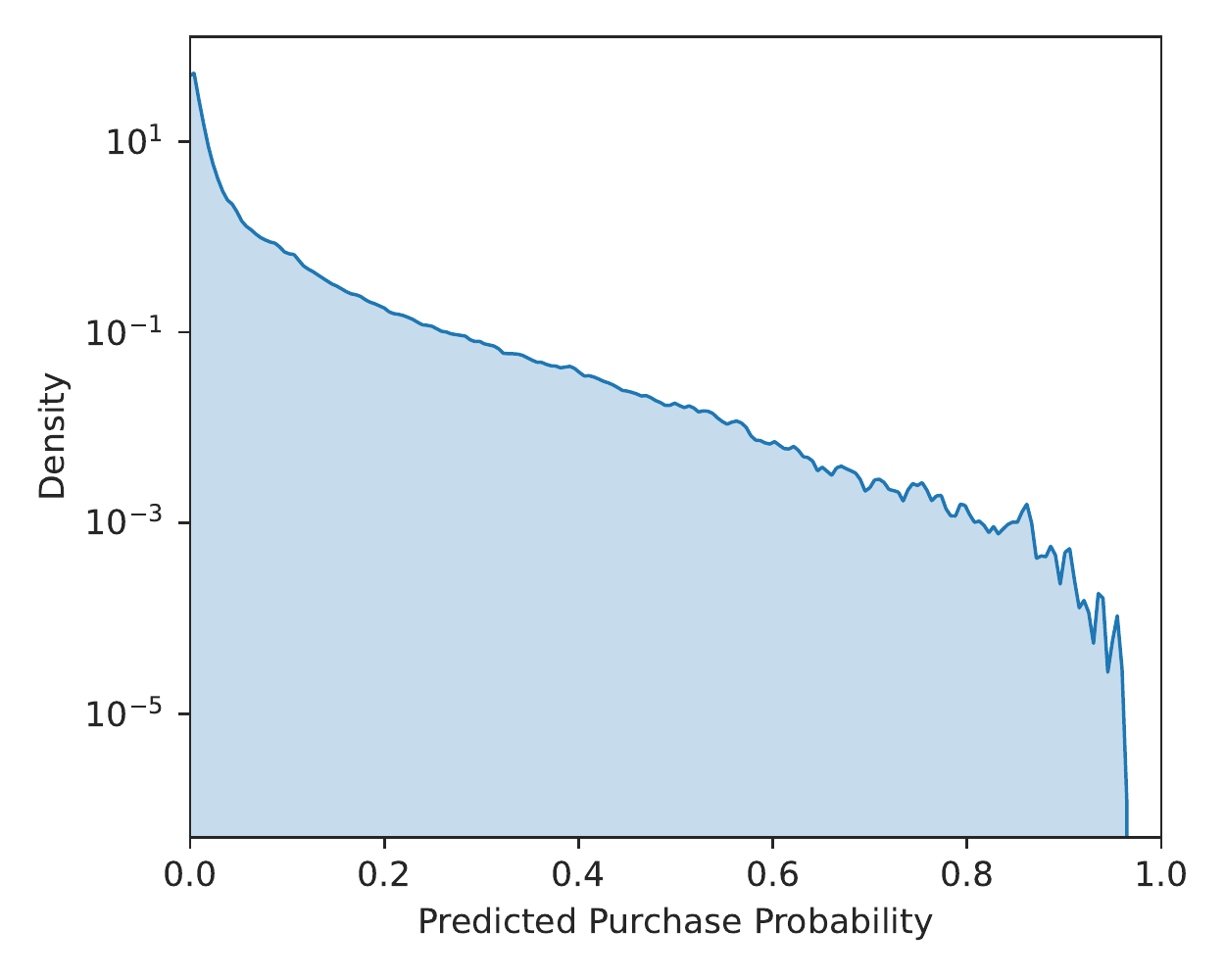}
	}
	\subfigure[LTV scores (x-axis is in log-scale).]{
	\includegraphics[width=0.47\linewidth]{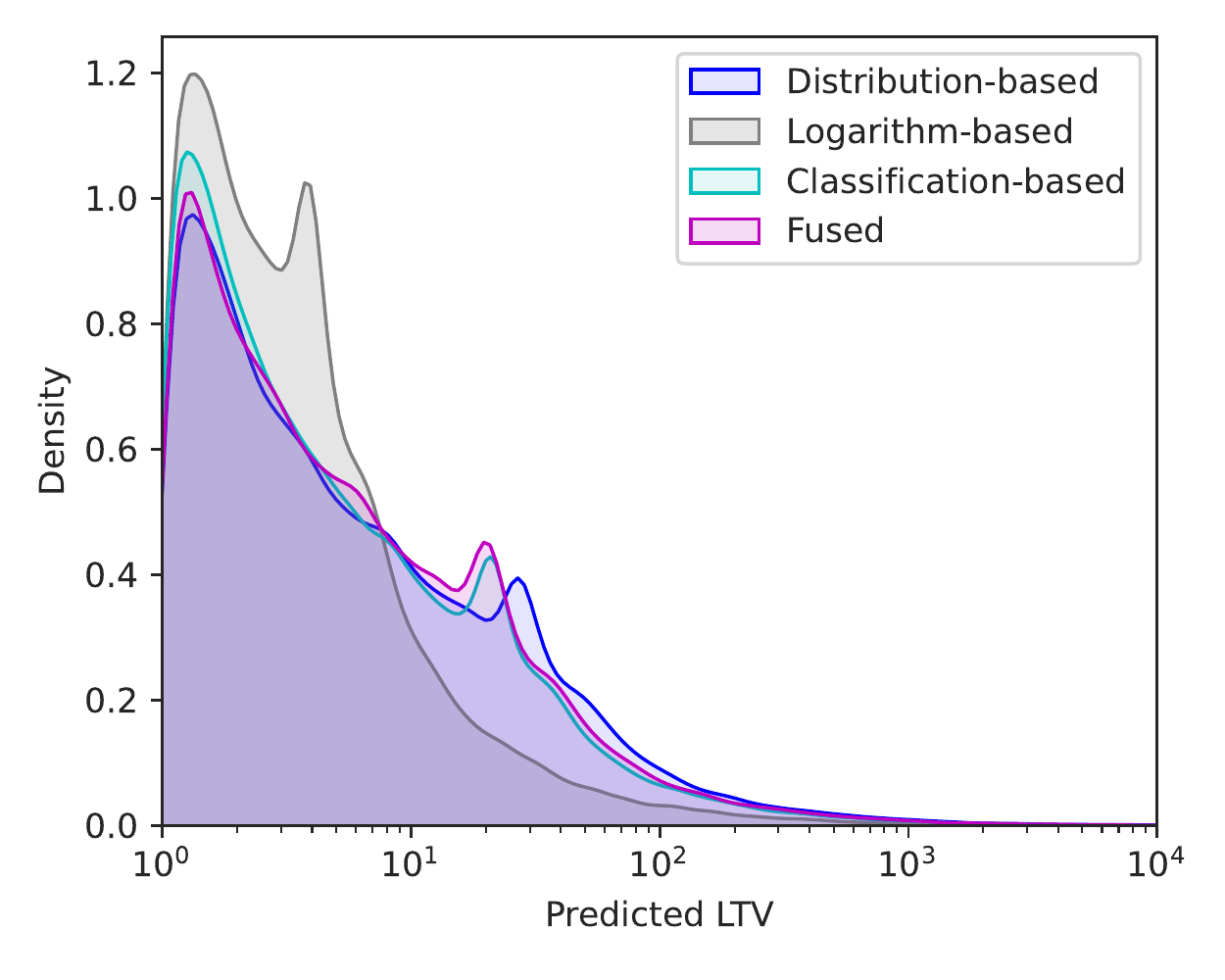}
	}
\caption{The distribution of predicted purchase probabilities and LTV scores.}\label{fig.ex6}
\end{figure}

We then compare our proposed contrastive learning approach with other mature contrastive learning strategies, including the standard pair-wise method and the data augmentation-based method introduced by~\cite{wang2022cl4ctr}.
From the results shown in Fig.~\ref{fig.ex5}, we find that the direct pair-wise contrast mechanism has a negative effect on model performance.
This is intuitive because it may suffer from heavy data noise and unwanted prediction bias brought by the contrast.
The augmentation-based contrast method is also inferior to our hybrid contrast method.
This is because the former can only enhance the discriminativeness of hidden representations and is agnostic to the relatedness between real-valued labels of different samples.
Our hybrid method can collaboratively exploit the supervision signals across different samples in both classification and regression tasks, which is especially powerful in LTV prediction.

\begin{figure}[!t]
	\centering
\subfigure[All samples, linear scale.]{
	\includegraphics[width=0.4\linewidth]{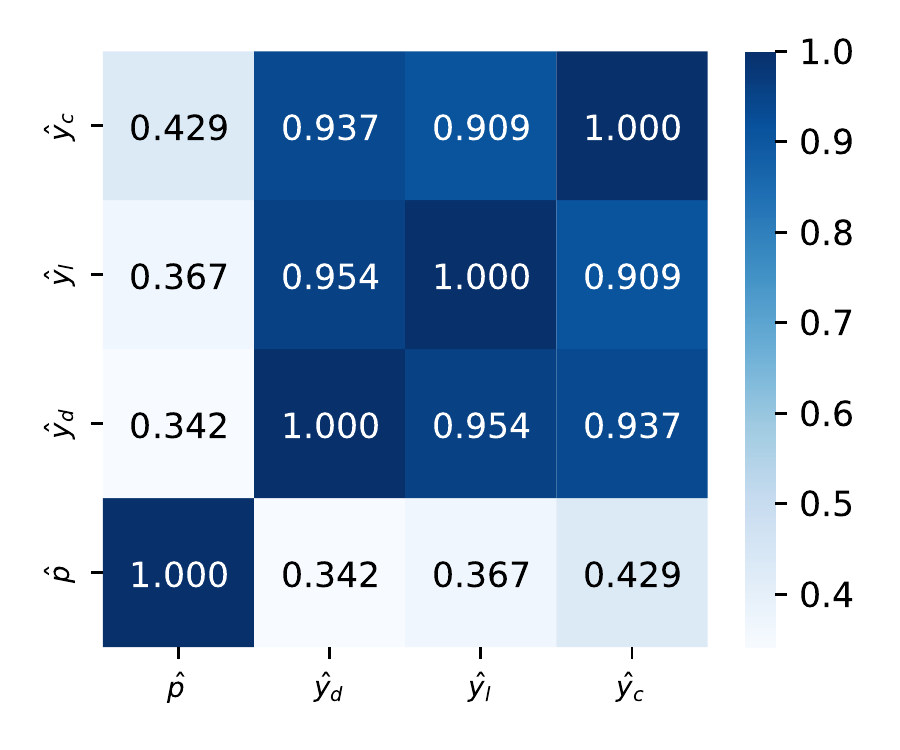}
	}
	\subfigure[Positive samples, linear scale.]{
	\includegraphics[width=0.4\linewidth]{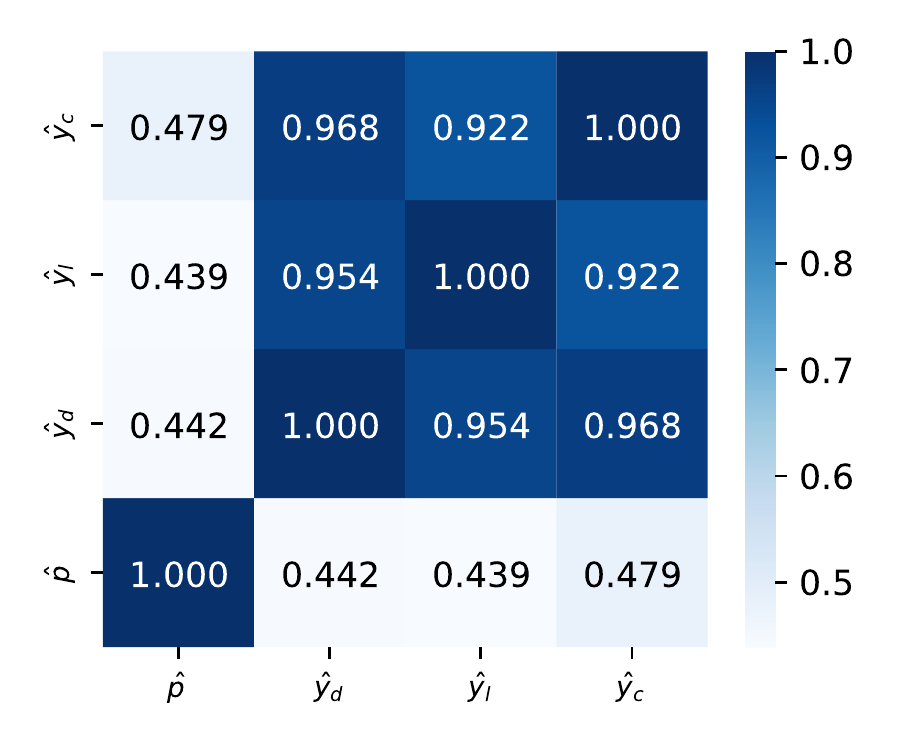}
	}
		\subfigure[All samples, log scale.]{
	\includegraphics[width=0.4\linewidth]{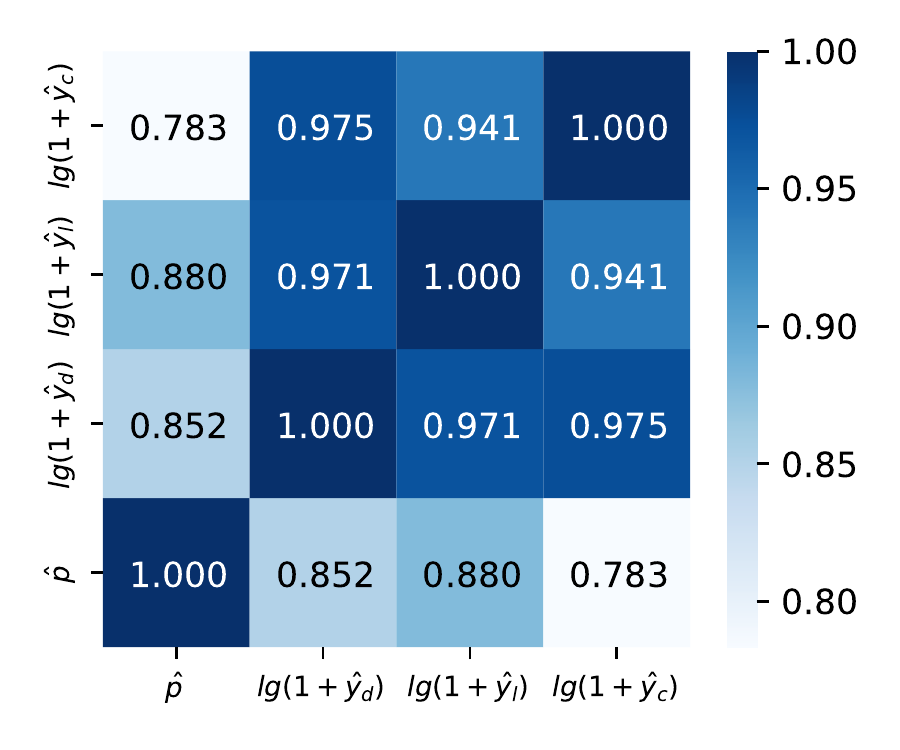}
	}
		\subfigure[Positive samples, log scale.]{
	\includegraphics[width=0.4\linewidth]{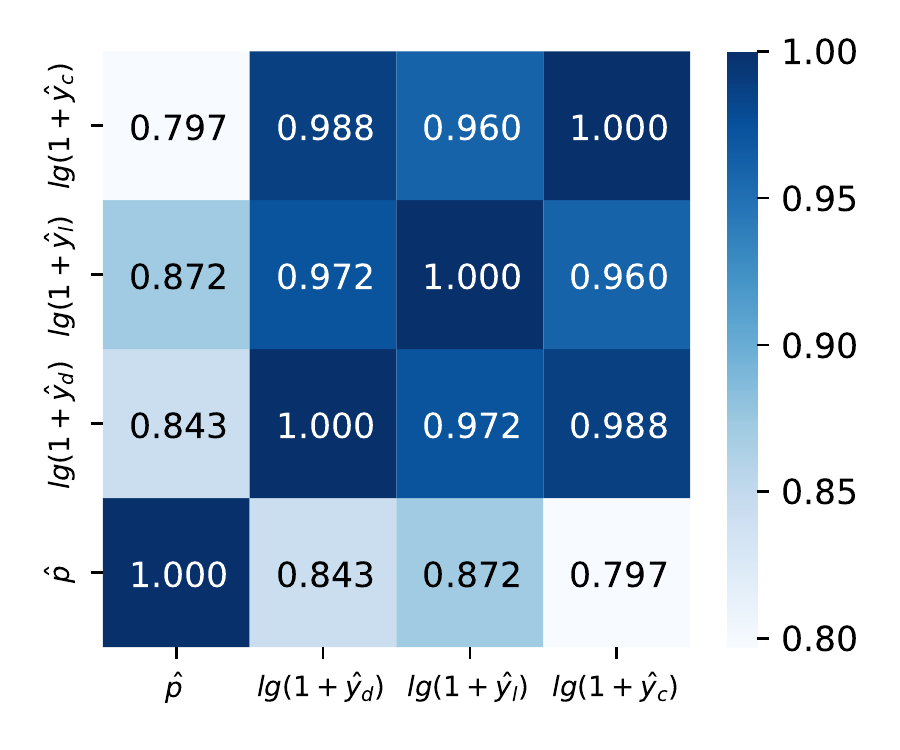}
	}
\caption{The Pearson correlation coefficients between different types of scores in linear/log scales computed on all samples or positive samples only.}\label{fig.ex7}
\end{figure}

\begin{figure*}[!t]
	\centering
	\includegraphics[width=0.999\linewidth]{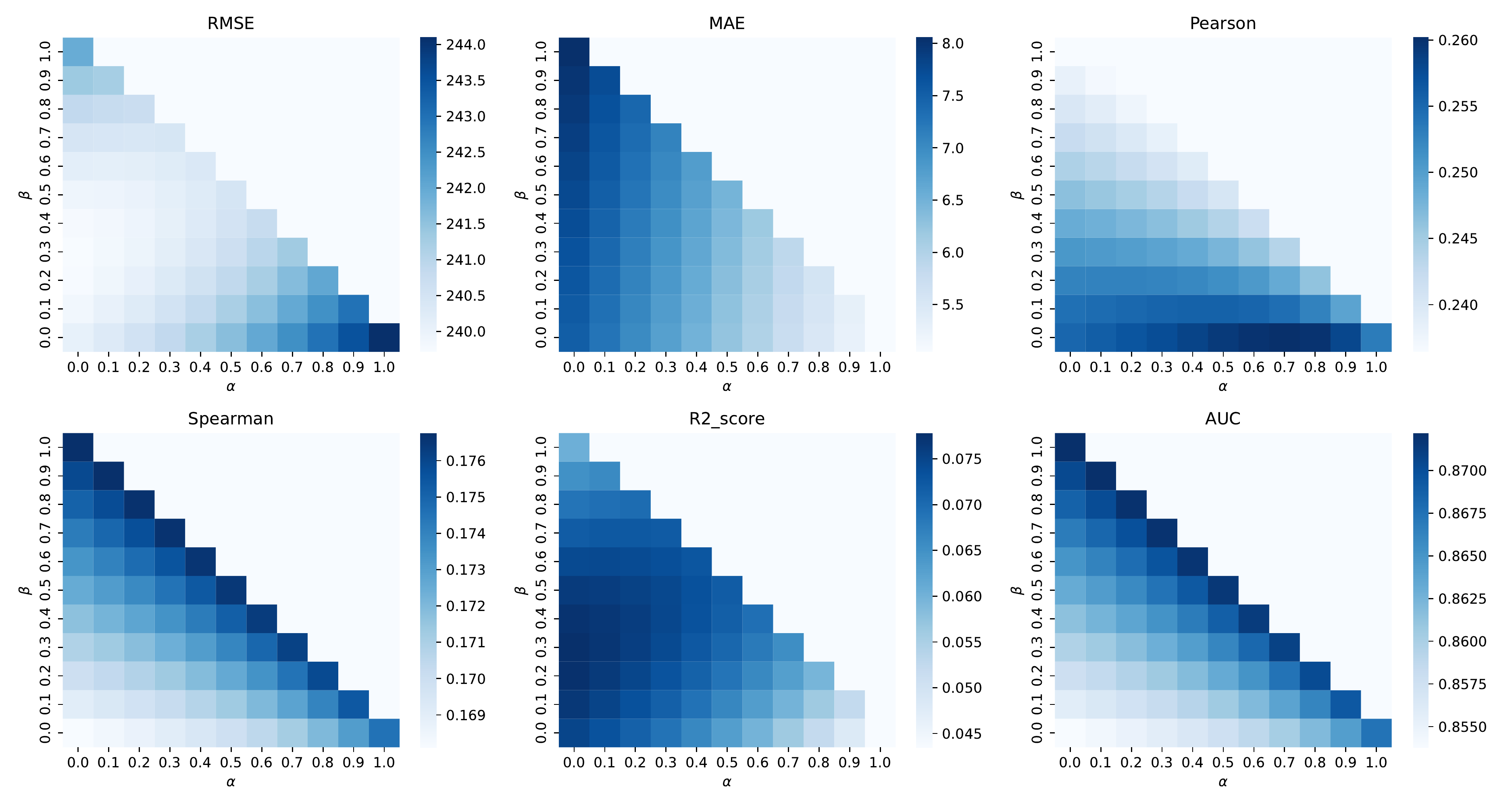}
\caption{Influence of the prediction fusing coefficients on final performance.}\label{fig.ex8}
\end{figure*}

\subsection{Prediction Results Analysis}

We then present some analysis of scores predicted by \textit{CMLTV}.
We first illustrate the distributions of purchase probabilities and the predicted LTV scores in Fig.~\ref{fig.ex6}.
Note that the LTV predictions that are smaller than 1 are omitted. 
We find the predicted purchase probabilities of most samples are low.
This is intuitive because purchase behaviors are usually sparse.
In addition, we find the regression results given by different regressors have some differences in their distributions, while their shapes are generally long-tailed.
Thus, by fusing the predictions from multiple heterogeneous regressors, the model can have a more comprehensive and less biased understanding of data distribution.

We also conduct further analysis of the prediction scores output by our method.
We compute the Pearson correlation coefficients between different types of scores on all samples or positive samples only, where the LTV regression scores are presented in both linear and log scales, as shown in Fig.~\ref{fig.ex7}.
We find that the regression results generated by different regressors are highly correlated (especially on positive samples) but not identical.
This result shows that different regression modules have high agreements on most samples but still keep their characteristics.
Therefore, it is safe to synthesize the predictions given by the three regressors.
In addition, the predicted purchase probabilities show strong correlations with predicted LTV scores, especially in the log scale.
This is mainly due to the regularization of  our contrastive regression loss so that predicted purchase probabilities are positively correlated with LTV regressions.

\subsection{Hyperparameter Analysis}

In this section we study the influence of two key hyperparameters on our approach, i.e., the prediction fusing coefficients $\alpha$ and $\beta$, which controls the relative importance of distribution-based and logarithm-based regressors in the final predictions.
We vary the values of both coefficients and the corresponding results are shown in Fig.~\ref{fig.ex8}.
From the results, we find the optimal points on the heatmaps of different metrics are diverse.
For example, the best MAE is achieved when $\alpha=1$, while the best Spearman score is achieved when $\alpha=\beta=0$.
Nonetheless, to achieve a good tradeoff between different metrics, we prefer to select moderate values for these coefficients.
In our experiments we set $\alpha=\beta=0.3$, which ensures that all three types of regressors can effectively contribute to the final prediction meanwhile all types of metrics are satisfactory at this point.
In real-world scenarios, we recommend practitioners search these hyperparameters according to the key metrics in the corresponding applications.

\begin{figure}[!t]
	\centering
	\includegraphics[width=0.8\linewidth]{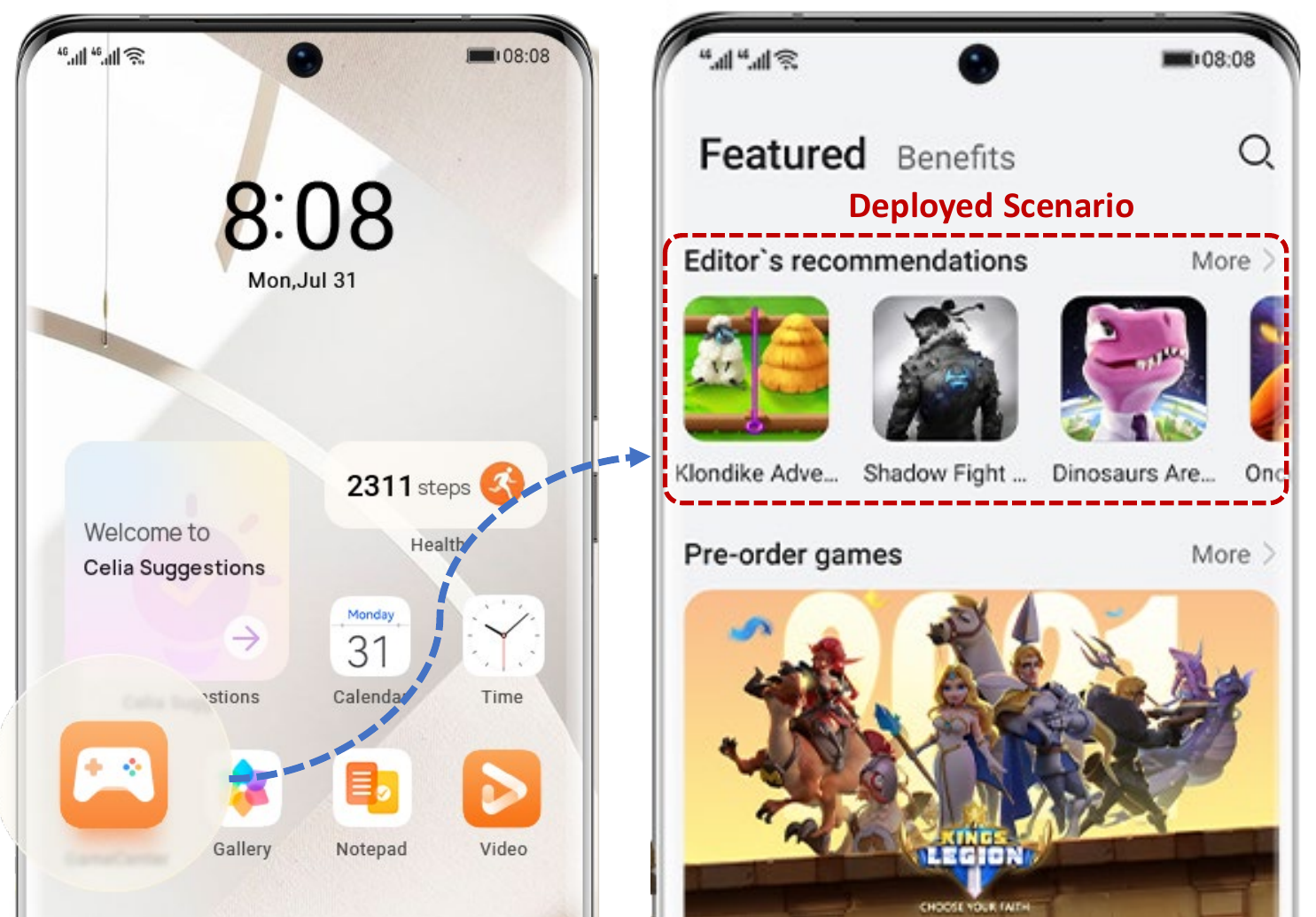}
\caption{The application scenario of our solution.}\label{fig.deploy}
\end{figure}

\section{Online Evaluation}

We deployed our \textit{CMLTV} method in Huawei's mobile game center for game player LTV prediction.
The predictions given by our model are used to generate personalized game suggestions on the ``editor's recommendations'' game display board, which is shown in Fig.~\ref{fig.deploy}.
We performed a 70-day online A/B test on our platform to verify the effectiveness of our algorithm.

\subsection{Online Experimental Settings}

The online A/B test lasts 70 days from Nov. 11, 2022, to Jan. 20, 2023.
The competitor of our model is a well-crafted baseline model, which is also learned in an end-to-end manner via a multi-task learning framework.
For fair comparison, the baseline model and our approach share the same input features and basic model backbones.
Both models are retrained periodically on a dataset aggregated from the transactions in the past month.
Our model is deployed in a virtual Linux computing node with an NVIDIA Tesla V100 GPU (32GB of memory).
In the online A/B test, 20\% of randomly selected users are used as the experimental group while  another 20\% of randomly selected users are reserved as the control group.
The traffics from the experimental and control groups are served by our method and the baseline model, respectively.
The predicted LTVs are multiplied with the estimated CVR (in terms of App download) as the ranking criteria to generate the final recommendation lists.

\subsection{Online Results}

Here we introduce the results of online A/B test.
The traffics served by our model achieved 32.26\% payment amount improvements over the control group traffics with a cost of 9\% additional inference latency, which is fully tolerable in our system.\footnote{The scale of payment amount increase is relatively larger than other types of common metrics such as CTR and CVR. This is mainly because high consumption users could significantly boost the total payment amount.}
In our further analysis, this significant improvement is mainly due to the increased payment amount of users at low and medium consumption levels (the revenue contributed by high consumption users is similar for baseline and our approach).
This finding indicates a critical advantage of our approach in boosting the activeness of a broad range of users, rather than only targeting high-spending users.

Due to the excellent performance of our \textit{CMLTV} solution, it has become a core module in the user-centric ecosystem of our App store, serving the main traffic of hundreds of millions of users with various types of mobile devices.
Since both the multi-view regression and hybrid contrastive learning modules in our approach are plug-and-play techniques, our solution has the potential to empower various LTV prediction models in different scenarios.
Compared with many existing solutions, our approach has minimal requirements for the data format and model pipeline, thereby can be adopted by different industrial systems with light efforts, and we are currently working on transferring our solution to other products such as online education and advertising.

%% file: data/conclusion.tex
\section{Conclusions and Limitations}\label{sec:Conclusion}

In this paper, we present an industrial solution to customer lifetime value prediction, named \textit{CMLTV}.
Different from conventional LTV prediction paradigms, we propose a multi-view regression framework that assembles multiple heterogeneous regression networks to capture complementary knowledge, which endows the model with better prediction accuracy and robustness when handling noisy and volatile consumption data.
In addition, we propose a hybrid contrastive learning mechanism to explicitly encode the relation information between samples into model learning meanwhile building organic connections between the binary purchase classifier and LTV regressors in our model.
Hence, the potential of training samples can be better exploited to help combat data sparsity.
Extensive experiments in both offline and online environments validate the superiority of our approach over baseline methods.
We have deployed our solution online to serve hundreds of millions of our mobile users and improve their experience.

Despite the effectiveness of our method, it still has the following limitations.
First, the model may underestimate the real LTV of some users since the predicted purchase probabilities are usually always smaller than one.
Second, it is difficult for our method to predict very high LTVs, since these outlier samples are very sparse in the training data.
Third, our hybrid contrastive learning method requires a large batch size, otherwise single batch has too few positive samples and the loss is volatile.
Finally, the LTVs for cold-start users and items are hard to predict due to the lack of prior knowledge~\cite{singh2009generalized}.
We will explore these directions in the future.

%% file: main.bbl

\begin{thebibliography}{54}


\ifx \showCODEN    \undefined \def \showCODEN     #1{\unskip}     \fi
\ifx \showDOI      \undefined \def \showDOI       #1{#1}\fi
\ifx \showISBNx    \undefined \def \showISBNx     #1{\unskip}     \fi
\ifx \showISBNxiii \undefined \def \showISBNxiii  #1{\unskip}     \fi
\ifx \showISSN     \undefined \def \showISSN      #1{\unskip}     \fi
\ifx \showLCCN     \undefined \def \showLCCN      #1{\unskip}     \fi
\ifx \shownote     \undefined \def \shownote      #1{#1}          \fi
\ifx \showarticletitle \undefined \def \showarticletitle #1{#1}   \fi
\ifx \showURL      \undefined \def \showURL       {\relax}        \fi
\providecommand\bibfield[2]{#2}
\providecommand\bibinfo[2]{#2}
\providecommand\natexlab[1]{#1}
\providecommand\showeprint[2][]{arXiv:#2}

\bibitem[\protect\citeauthoryear{Akram, Hui, Khan, Tanveer, Mehmood, and
  Ahmad}{Akram et~al\mbox{.}}{2018}]%
        {akram2018website}
\bibfield{author}{\bibinfo{person}{Umair Akram}, \bibinfo{person}{Peng Hui},
  \bibinfo{person}{Muhammad~Kaleem Khan}, \bibinfo{person}{Yasir Tanveer},
  \bibinfo{person}{Khalid Mehmood}, {and} \bibinfo{person}{Wasim Ahmad}.}
  \bibinfo{year}{2018}\natexlab{}.
\newblock \showarticletitle{How website quality affects online impulse buying:
  Moderating effects of sales promotion and credit card use}.
\newblock \bibinfo{journal}{\emph{Asia Pacific Journal of Marketing and
  Logistics}} (\bibinfo{year}{2018}).
\newblock


\bibitem[\protect\citeauthoryear{Bai, Xiao, Wu, Yang, Yu, and Nie}{Bai
  et~al\mbox{.}}{2022}]%
        {bai2022contrastive}
\bibfield{author}{\bibinfo{person}{Ting Bai}, \bibinfo{person}{Yudong Xiao},
  \bibinfo{person}{Bin Wu}, \bibinfo{person}{Guojun Yang},
  \bibinfo{person}{Hongyong Yu}, {and} \bibinfo{person}{Jian-Yun Nie}.}
  \bibinfo{year}{2022}\natexlab{}.
\newblock \showarticletitle{A Contrastive Sharing Model for Multi-Task
  Recommendation}. In \bibinfo{booktitle}{\emph{WWW}}.
  \bibinfo{pages}{3239--3247}.
\newblock


\bibitem[\protect\citeauthoryear{Bauer and Jannach}{Bauer and Jannach}{2021}]%
        {bauer2021improved}
\bibfield{author}{\bibinfo{person}{Josef Bauer} {and} \bibinfo{person}{Dietmar
  Jannach}.} \bibinfo{year}{2021}\natexlab{}.
\newblock \showarticletitle{Improved Customer Lifetime Value Prediction With
  Sequence-To-Sequence Learning and Feature-Based Models}.
\newblock \bibinfo{journal}{\emph{TKDD}} \bibinfo{volume}{15},
  \bibinfo{number}{5} (\bibinfo{year}{2021}), \bibinfo{pages}{1--37}.
\newblock


\bibitem[\protect\citeauthoryear{Bengio and LeCun}{Bengio and LeCun}{2015}]%
        {kingma2014adam}
\bibfield{author}{\bibinfo{person}{Yoshua Bengio} {and} \bibinfo{person}{Yann
  LeCun}.} \bibinfo{year}{2015}\natexlab{}.
\newblock \showarticletitle{Adam: {A} Method for Stochastic Optimization}. In
  \bibinfo{booktitle}{\emph{ICLR}}.
\newblock


\bibitem[\protect\citeauthoryear{Berger and Nasr}{Berger and Nasr}{1998}]%
        {berger1998customer}
\bibfield{author}{\bibinfo{person}{Paul~D Berger} {and} \bibinfo{person}{Nada~I
  Nasr}.} \bibinfo{year}{1998}\natexlab{}.
\newblock \showarticletitle{Customer lifetime value: Marketing models and
  applications}.
\newblock \bibinfo{journal}{\emph{Journal of interactive marketing}}
  \bibinfo{volume}{12}, \bibinfo{number}{1} (\bibinfo{year}{1998}),
  \bibinfo{pages}{17--30}.
\newblock


\bibitem[\protect\citeauthoryear{Blattberg, Malthouse, and Neslin}{Blattberg
  et~al\mbox{.}}{2009}]%
        {blattberg2009customer}
\bibfield{author}{\bibinfo{person}{Robert~C Blattberg},
  \bibinfo{person}{Edward~C Malthouse}, {and} \bibinfo{person}{Scott~A
  Neslin}.} \bibinfo{year}{2009}\natexlab{}.
\newblock \showarticletitle{Customer lifetime value: Empirical generalizations
  and some conceptual questions}.
\newblock \bibinfo{journal}{\emph{Journal of Interactive Marketing}}
  \bibinfo{volume}{23}, \bibinfo{number}{2} (\bibinfo{year}{2009}),
  \bibinfo{pages}{157--168}.
\newblock


\bibitem[\protect\citeauthoryear{Borle, Singh, and Jain}{Borle
  et~al\mbox{.}}{2008}]%
        {borle2008customer}
\bibfield{author}{\bibinfo{person}{Sharad Borle}, \bibinfo{person}{Siddharth~S
  Singh}, {and} \bibinfo{person}{Dipak~C Jain}.}
  \bibinfo{year}{2008}\natexlab{}.
\newblock \showarticletitle{Customer lifetime value measurement}.
\newblock \bibinfo{journal}{\emph{Management science}} \bibinfo{volume}{54},
  \bibinfo{number}{1} (\bibinfo{year}{2008}), \bibinfo{pages}{100--112}.
\newblock


\bibitem[\protect\citeauthoryear{Burelli}{Burelli}{2019}]%
        {burelli2019predicting}
\bibfield{author}{\bibinfo{person}{Paolo Burelli}.}
  \bibinfo{year}{2019}\natexlab{}.
\newblock \showarticletitle{Predicting customer lifetime value in free-to-play
  games}.
\newblock In \bibinfo{booktitle}{\emph{Data analytics applications in gaming
  and entertainment}}. \bibinfo{publisher}{Auerbach Publications},
  \bibinfo{pages}{79--107}.
\newblock


\bibitem[\protect\citeauthoryear{Chamberlain, Cardoso, Liu, Pagliari, and
  Deisenroth}{Chamberlain et~al\mbox{.}}{2017}]%
        {chamberlain2017customer}
\bibfield{author}{\bibinfo{person}{Benjamin~Paul Chamberlain},
  \bibinfo{person}{Angelo Cardoso}, \bibinfo{person}{CH~Bryan Liu},
  \bibinfo{person}{Roberto Pagliari}, {and} \bibinfo{person}{Marc~Peter
  Deisenroth}.} \bibinfo{year}{2017}\natexlab{}.
\newblock \showarticletitle{Customer lifetime value prediction using
  embeddings}. In \bibinfo{booktitle}{\emph{KDD}}. \bibinfo{pages}{1753--1762}.
\newblock


\bibitem[\protect\citeauthoryear{Chen, Guitart, del R{\'\i}o, and
  Peri{\'a}nez}{Chen et~al\mbox{.}}{2018}]%
        {chen2018customer}
\bibfield{author}{\bibinfo{person}{Pei~Pei Chen}, \bibinfo{person}{Anna
  Guitart}, \bibinfo{person}{Ana~Fern{\'a}ndez del R{\'\i}o}, {and}
  \bibinfo{person}{Africa Peri{\'a}nez}.} \bibinfo{year}{2018}\natexlab{}.
\newblock \showarticletitle{Customer lifetime value in video games using deep
  learning and parametric models}. In \bibinfo{booktitle}{\emph{Big Data}}.
  IEEE, \bibinfo{pages}{2134--2140}.
\newblock


\bibitem[\protect\citeauthoryear{Dahana, Miwa, and Morisada}{Dahana
  et~al\mbox{.}}{2019}]%
        {dahana2019linking}
\bibfield{author}{\bibinfo{person}{Wirawan~Dony Dahana},
  \bibinfo{person}{Yukihiro Miwa}, {and} \bibinfo{person}{Makoto Morisada}.}
  \bibinfo{year}{2019}\natexlab{}.
\newblock \showarticletitle{Linking lifestyle to customer lifetime value: An
  exploratory study in an online fashion retail market}.
\newblock \bibinfo{journal}{\emph{Journal of Business Research}}
  \bibinfo{volume}{99} (\bibinfo{year}{2019}), \bibinfo{pages}{319--331}.
\newblock


\bibitem[\protect\citeauthoryear{del R{\'\i}o, Chen, and Peri{\'a}nez}{del
  R{\'\i}o et~al\mbox{.}}{2019}]%
        {del2019profiling}
\bibfield{author}{\bibinfo{person}{Ana~Fern{\'a}ndez del R{\'\i}o},
  \bibinfo{person}{Pei~Pei Chen}, {and} \bibinfo{person}{Africa Peri{\'a}nez}.}
  \bibinfo{year}{2019}\natexlab{}.
\newblock \showarticletitle{Profiling players with engagement predictions}. In
  \bibinfo{booktitle}{\emph{CoG}}. IEEE, \bibinfo{pages}{1--4}.
\newblock


\bibitem[\protect\citeauthoryear{Drachen, Pastor, Liu, Fontaine, Chang, Runge,
  Sifa, and Klabjan}{Drachen et~al\mbox{.}}{2018}]%
        {drachen2018to}
\bibfield{author}{\bibinfo{person}{Anders Drachen}, \bibinfo{person}{Mari
  Pastor}, \bibinfo{person}{Aron Liu}, \bibinfo{person}{Dylan~Jack Fontaine},
  \bibinfo{person}{Yuan Chang}, \bibinfo{person}{Julian Runge},
  \bibinfo{person}{Rafet Sifa}, {and} \bibinfo{person}{Diego Klabjan}.}
  \bibinfo{year}{2018}\natexlab{}.
\newblock \showarticletitle{To be or not to be...social: incorporating simple
  social features in mobile game customer lifetime value predictions}. In
  \bibinfo{booktitle}{\emph{ACSW}}. \bibinfo{publisher}{{ACM}},
  \bibinfo{pages}{40:1--40:10}.
\newblock


\bibitem[\protect\citeauthoryear{Fader, Hardie, and Lee}{Fader
  et~al\mbox{.}}{2005}]%
        {fader2005rfm}
\bibfield{author}{\bibinfo{person}{Peter~S Fader}, \bibinfo{person}{Bruce~GS
  Hardie}, {and} \bibinfo{person}{Ka~Lok Lee}.}
  \bibinfo{year}{2005}\natexlab{}.
\newblock \showarticletitle{RFM and CLV: Using iso-value curves for customer
  base analysis}.
\newblock \bibinfo{journal}{\emph{Journal of marketing research}}
  \bibinfo{volume}{42}, \bibinfo{number}{4} (\bibinfo{year}{2005}),
  \bibinfo{pages}{415--430}.
\newblock


\bibitem[\protect\citeauthoryear{Glady, Baesens, and Croux}{Glady
  et~al\mbox{.}}{2009a}]%
        {glady2009modeling}
\bibfield{author}{\bibinfo{person}{Nicolas Glady}, \bibinfo{person}{Bart
  Baesens}, {and} \bibinfo{person}{Christophe Croux}.}
  \bibinfo{year}{2009}\natexlab{a}.
\newblock \showarticletitle{Modeling churn using customer lifetime value}.
\newblock \bibinfo{journal}{\emph{European Journal of Operational Research}}
  \bibinfo{volume}{197}, \bibinfo{number}{1} (\bibinfo{year}{2009}),
  \bibinfo{pages}{402--411}.
\newblock


\bibitem[\protect\citeauthoryear{Glady, Baesens, and Croux}{Glady
  et~al\mbox{.}}{2009b}]%
        {glady2009modified}
\bibfield{author}{\bibinfo{person}{Nicolas Glady}, \bibinfo{person}{Bart
  Baesens}, {and} \bibinfo{person}{Christophe Croux}.}
  \bibinfo{year}{2009}\natexlab{b}.
\newblock \showarticletitle{A modified Pareto/NBD approach for predicting
  customer lifetime value}.
\newblock \bibinfo{journal}{\emph{Expert Systems with Applications}}
  \bibinfo{volume}{36}, \bibinfo{number}{2} (\bibinfo{year}{2009}),
  \bibinfo{pages}{2062--2071}.
\newblock


\bibitem[\protect\citeauthoryear{Guo, Tang, Ye, Li, and He}{Guo
  et~al\mbox{.}}{2017}]%
        {guo2017deepfm}
\bibfield{author}{\bibinfo{person}{Huifeng Guo}, \bibinfo{person}{Ruiming
  Tang}, \bibinfo{person}{Yunming Ye}, \bibinfo{person}{Zhenguo Li}, {and}
  \bibinfo{person}{Xiuqiang He}.} \bibinfo{year}{2017}\natexlab{}.
\newblock \showarticletitle{DeepFM: a factorization-machine based neural
  network for CTR prediction}. In \bibinfo{booktitle}{\emph{IJCAI}}.
  \bibinfo{pages}{1725--1731}.
\newblock


\bibitem[\protect\citeauthoryear{Gupta, Hanssens, Hardie, Kahn, Kumar, Lin,
  Ravishanker, and Sriram}{Gupta et~al\mbox{.}}{2006}]%
        {gupta2006modeling}
\bibfield{author}{\bibinfo{person}{Sunil Gupta}, \bibinfo{person}{Dominique
  Hanssens}, \bibinfo{person}{Bruce Hardie}, \bibinfo{person}{Wiliam Kahn},
  \bibinfo{person}{V Kumar}, \bibinfo{person}{Nathaniel Lin},
  \bibinfo{person}{Nalini Ravishanker}, {and} \bibinfo{person}{S Sriram}.}
  \bibinfo{year}{2006}\natexlab{}.
\newblock \showarticletitle{Modeling customer lifetime value}.
\newblock \bibinfo{journal}{\emph{Journal of service research}}
  \bibinfo{volume}{9}, \bibinfo{number}{2} (\bibinfo{year}{2006}),
  \bibinfo{pages}{139--155}.
\newblock


\bibitem[\protect\citeauthoryear{Ioffe and Szegedy}{Ioffe and Szegedy}{2015}]%
        {ioffe2015batch}
\bibfield{author}{\bibinfo{person}{Sergey Ioffe} {and}
  \bibinfo{person}{Christian Szegedy}.} \bibinfo{year}{2015}\natexlab{}.
\newblock \showarticletitle{Batch normalization: Accelerating deep network
  training by reducing internal covariate shift}. In
  \bibinfo{booktitle}{\emph{ICML}}. pmlr, \bibinfo{pages}{448--456}.
\newblock


\bibitem[\protect\citeauthoryear{Jain and Singh}{Jain and Singh}{2002}]%
        {jain2002customer}
\bibfield{author}{\bibinfo{person}{Dipak Jain} {and}
  \bibinfo{person}{Siddhartha~S Singh}.} \bibinfo{year}{2002}\natexlab{}.
\newblock \showarticletitle{Customer lifetime value research in marketing: A
  review and future directions}.
\newblock \bibinfo{journal}{\emph{Journal of interactive marketing}}
  \bibinfo{volume}{16}, \bibinfo{number}{2} (\bibinfo{year}{2002}),
  \bibinfo{pages}{34--46}.
\newblock


\bibitem[\protect\citeauthoryear{Jasek, Vrana, Sperkova, Smutny, and
  Kobulsky}{Jasek et~al\mbox{.}}{2019}]%
        {jasek2019comparative}
\bibfield{author}{\bibinfo{person}{Pavel Jasek}, \bibinfo{person}{Lenka Vrana},
  \bibinfo{person}{Lucie Sperkova}, \bibinfo{person}{Zdenek Smutny}, {and}
  \bibinfo{person}{Marek Kobulsky}.} \bibinfo{year}{2019}\natexlab{}.
\newblock \showarticletitle{Comparative analysis of selected probabilistic
  customer lifetime value models in online shopping}.
\newblock \bibinfo{journal}{\emph{Journal of Business Economics and
  Management}} \bibinfo{volume}{20}, \bibinfo{number}{3}
  (\bibinfo{year}{2019}), \bibinfo{pages}{398--423}.
\newblock


\bibitem[\protect\citeauthoryear{Jiang}{Jiang}{2020}]%
        {jiang2020study}
\bibfield{author}{\bibinfo{person}{Junxiang Jiang}.}
  \bibinfo{year}{2020}\natexlab{}.
\newblock \showarticletitle{A Study of Game Payment Data Mining: Predicting
  High-Value Users for MMORPGs}. In \bibinfo{booktitle}{\emph{PAKDD
  Workshops}}. Springer, \bibinfo{pages}{181--192}.
\newblock


\bibitem[\protect\citeauthoryear{Kim and Yum}{Kim and Yum}{2008}]%
        {kim2008selection}
\bibfield{author}{\bibinfo{person}{Jin~Seon Kim} {and}
  \bibinfo{person}{Bong-Jin Yum}.} \bibinfo{year}{2008}\natexlab{}.
\newblock \showarticletitle{Selection between Weibull and lognormal
  distributions: A comparative simulation study}.
\newblock \bibinfo{journal}{\emph{Computational Statistics \& Data Analysis}}
  \bibinfo{volume}{53}, \bibinfo{number}{2} (\bibinfo{year}{2008}),
  \bibinfo{pages}{477--485}.
\newblock


\bibitem[\protect\citeauthoryear{Kumar, Ramani, and Bohling}{Kumar
  et~al\mbox{.}}{2004}]%
        {kumar2004customer}
\bibfield{author}{\bibinfo{person}{V Kumar}, \bibinfo{person}{Girish Ramani},
  {and} \bibinfo{person}{Timothy Bohling}.} \bibinfo{year}{2004}\natexlab{}.
\newblock \showarticletitle{Customer lifetime value approaches and best
  practice applications}.
\newblock \bibinfo{journal}{\emph{Journal of interactive Marketing}}
  \bibinfo{volume}{18}, \bibinfo{number}{3} (\bibinfo{year}{2004}),
  \bibinfo{pages}{60--72}.
\newblock


\bibitem[\protect\citeauthoryear{Li, Shao, Yang, Fang, and Song}{Li
  et~al\mbox{.}}{2022}]%
        {li2022billion}
\bibfield{author}{\bibinfo{person}{Kunpeng Li}, \bibinfo{person}{Guangcui
  Shao}, \bibinfo{person}{Naijun Yang}, \bibinfo{person}{Xiao Fang}, {and}
  \bibinfo{person}{Yang Song}.} \bibinfo{year}{2022}\natexlab{}.
\newblock \showarticletitle{Billion-user Customer Lifetime Value Prediction: An
  Industrial-scale Solution from Kuaishou}. In
  \bibinfo{booktitle}{\emph{CIKM}}. \bibinfo{publisher}{{ACM}},
  \bibinfo{pages}{3243--3251}.
\newblock


\bibitem[\protect\citeauthoryear{Lin, Yang, Liu, Peng, Zhao, Wang, and
  Zheng}{Lin et~al\mbox{.}}{2022}]%
        {lin2022personalized}
\bibfield{author}{\bibinfo{person}{Zihan Lin}, \bibinfo{person}{Xuanhua Yang},
  \bibinfo{person}{Shaoguo Liu}, \bibinfo{person}{Xiaoyu Peng},
  \bibinfo{person}{Wayne~Xin Zhao}, \bibinfo{person}{Liang Wang}, {and}
  \bibinfo{person}{Bo Zheng}.} \bibinfo{year}{2022}\natexlab{}.
\newblock \showarticletitle{Personalized Inter-Task Contrastive Learning for
  CTR\&CVR Joint Estimation}.
\newblock \bibinfo{journal}{\emph{arXiv preprint arXiv:2208.13442}}
  (\bibinfo{year}{2022}).
\newblock


\bibitem[\protect\citeauthoryear{Malthouse and Blattberg}{Malthouse and
  Blattberg}{2005}]%
        {malthouse2005can}
\bibfield{author}{\bibinfo{person}{Edward~C Malthouse} {and}
  \bibinfo{person}{Robert~C Blattberg}.} \bibinfo{year}{2005}\natexlab{}.
\newblock \showarticletitle{Can we predict customer lifetime value?}
\newblock \bibinfo{journal}{\emph{Journal of interactive marketing}}
  \bibinfo{volume}{19}, \bibinfo{number}{1} (\bibinfo{year}{2005}),
  \bibinfo{pages}{2--16}.
\newblock


\bibitem[\protect\citeauthoryear{Mani, Drew, Betz, and Datta}{Mani
  et~al\mbox{.}}{1999}]%
        {mani1999statistics}
\bibfield{author}{\bibinfo{person}{DR Mani}, \bibinfo{person}{James Drew},
  \bibinfo{person}{Andrew Betz}, {and} \bibinfo{person}{Piew Datta}.}
  \bibinfo{year}{1999}\natexlab{}.
\newblock \showarticletitle{Statistics and data mining techniques for lifetime
  value modeling}. In \bibinfo{booktitle}{\emph{KDD}}.
  \bibinfo{pages}{94--103}.
\newblock


\bibitem[\protect\citeauthoryear{Oord, Li, and Vinyals}{Oord
  et~al\mbox{.}}{2018}]%
        {oord2018representation}
\bibfield{author}{\bibinfo{person}{Aaron van~den Oord}, \bibinfo{person}{Yazhe
  Li}, {and} \bibinfo{person}{Oriol Vinyals}.} \bibinfo{year}{2018}\natexlab{}.
\newblock \showarticletitle{Representation learning with contrastive predictive
  coding}.
\newblock \bibinfo{journal}{\emph{arXiv preprint arXiv:1807.03748}}
  (\bibinfo{year}{2018}).
\newblock


\bibitem[\protect\citeauthoryear{Pan, Yao, Han, Jia, Zhang, and Yang}{Pan
  et~al\mbox{.}}{2021}]%
        {pan2021click}
\bibfield{author}{\bibinfo{person}{Yujie Pan}, \bibinfo{person}{Jiangchao Yao},
  \bibinfo{person}{Bo Han}, \bibinfo{person}{Kunyang Jia}, \bibinfo{person}{Ya
  Zhang}, {and} \bibinfo{person}{Hongxia Yang}.}
  \bibinfo{year}{2021}\natexlab{}.
\newblock \showarticletitle{Click-through Rate Prediction with Auto-Quantized
  Contrastive Learning}.
\newblock \bibinfo{journal}{\emph{arXiv preprint arXiv:2109.13921}}
  (\bibinfo{year}{2021}).
\newblock


\bibitem[\protect\citeauthoryear{Pollak}{Pollak}{2021}]%
        {pollak2021predicting}
\bibfield{author}{\bibinfo{person}{Ziv Pollak}.}
  \bibinfo{year}{2021}\natexlab{}.
\newblock \showarticletitle{Predicting Customer Lifetime Values--ecommerce use
  case}.
\newblock \bibinfo{journal}{\emph{arXiv preprint arXiv:2102.05771}}
  (\bibinfo{year}{2021}).
\newblock


\bibitem[\protect\citeauthoryear{Rendle, Freudenthaler, Gantner, and
  Schmidt-Thieme}{Rendle et~al\mbox{.}}{2009}]%
        {rendle2009bpr}
\bibfield{author}{\bibinfo{person}{Steffen Rendle}, \bibinfo{person}{Christoph
  Freudenthaler}, \bibinfo{person}{Zeno Gantner}, {and} \bibinfo{person}{Lars
  Schmidt-Thieme}.} \bibinfo{year}{2009}\natexlab{}.
\newblock \showarticletitle{BPR: Bayesian personalized ranking from implicit
  feedback}. In \bibinfo{booktitle}{\emph{UAI}}. \bibinfo{pages}{452--461}.
\newblock


\bibitem[\protect\citeauthoryear{Rosset, Neumann, Eick, and Vatnik}{Rosset
  et~al\mbox{.}}{2003}]%
        {rosset2003customer}
\bibfield{author}{\bibinfo{person}{Saharon Rosset}, \bibinfo{person}{Einat
  Neumann}, \bibinfo{person}{Uri Eick}, {and} \bibinfo{person}{Nurit Vatnik}.}
  \bibinfo{year}{2003}\natexlab{}.
\newblock \showarticletitle{Customer lifetime value models for decision
  support}.
\newblock \bibinfo{journal}{\emph{Data mining and knowledge discovery}}
  \bibinfo{volume}{7} (\bibinfo{year}{2003}), \bibinfo{pages}{321--339}.
\newblock


\bibitem[\protect\citeauthoryear{Sabbeh}{Sabbeh}{2018}]%
        {sabbeh2018machine}
\bibfield{author}{\bibinfo{person}{Sahar~F Sabbeh}.}
  \bibinfo{year}{2018}\natexlab{}.
\newblock \showarticletitle{Machine-learning techniques for customer retention:
  A comparative study}.
\newblock \bibinfo{journal}{\emph{IJACSA}} \bibinfo{volume}{9},
  \bibinfo{number}{2} (\bibinfo{year}{2018}).
\newblock


\bibitem[\protect\citeauthoryear{Schmittlein, Morrison, and
  Colombo}{Schmittlein et~al\mbox{.}}{1987}]%
        {schmittlein1987counting}
\bibfield{author}{\bibinfo{person}{David~C Schmittlein},
  \bibinfo{person}{Donald~G Morrison}, {and} \bibinfo{person}{Richard
  Colombo}.} \bibinfo{year}{1987}\natexlab{}.
\newblock \showarticletitle{Counting your customers: Who-are they and what will
  they do next?}
\newblock \bibinfo{journal}{\emph{Management science}} \bibinfo{volume}{33},
  \bibinfo{number}{1} (\bibinfo{year}{1987}), \bibinfo{pages}{1--24}.
\newblock


\bibitem[\protect\citeauthoryear{Singh, Borle, and Jain}{Singh
  et~al\mbox{.}}{2009}]%
        {singh2009generalized}
\bibfield{author}{\bibinfo{person}{Siddharth~S Singh}, \bibinfo{person}{Sharad
  Borle}, {and} \bibinfo{person}{Dipak~C Jain}.}
  \bibinfo{year}{2009}\natexlab{}.
\newblock \showarticletitle{A generalized framework for estimating customer
  lifetime value when customer lifetimes are not observed}.
\newblock \bibinfo{journal}{\emph{Qme}}  \bibinfo{volume}{7}
  (\bibinfo{year}{2009}), \bibinfo{pages}{181--205}.
\newblock


\bibitem[\protect\citeauthoryear{Tekin, Kaya, and Cebi}{Tekin
  et~al\mbox{.}}{2022}]%
        {tekin2022customer}
\bibfield{author}{\bibinfo{person}{Ahmet~Tezcan Tekin}, \bibinfo{person}{Tolga
  Kaya}, {and} \bibinfo{person}{Ferhan Cebi}.} \bibinfo{year}{2022}\natexlab{}.
\newblock \showarticletitle{Customer lifetime value prediction for gaming
  industry: fuzzy clustering based approach}.
\newblock \bibinfo{journal}{\emph{Journal of Intelligent \& Fuzzy Systems}}
  \bibinfo{volume}{42}, \bibinfo{number}{1} (\bibinfo{year}{2022}),
  \bibinfo{pages}{87--96}.
\newblock


\bibitem[\protect\citeauthoryear{Tkachenko}{Tkachenko}{2015}]%
        {tkachenko2015autonomous}
\bibfield{author}{\bibinfo{person}{Yegor Tkachenko}.}
  \bibinfo{year}{2015}\natexlab{}.
\newblock \showarticletitle{Autonomous CRM control via CLV approximation with
  deep reinforcement learning in discrete and continuous action space}.
\newblock \bibinfo{journal}{\emph{arXiv preprint arXiv:1504.01840}}
  (\bibinfo{year}{2015}).
\newblock


\bibitem[\protect\citeauthoryear{Vanderveld, Pandey, Han, and
  Parekh}{Vanderveld et~al\mbox{.}}{2016}]%
        {vanderveld2016an}
\bibfield{author}{\bibinfo{person}{Ali Vanderveld}, \bibinfo{person}{Addhyan
  Pandey}, \bibinfo{person}{Angela Han}, {and} \bibinfo{person}{Rajesh
  Parekh}.} \bibinfo{year}{2016}\natexlab{}.
\newblock \showarticletitle{An Engagement-Based Customer Lifetime Value System
  for E-commerce}. In \bibinfo{booktitle}{\emph{KDD}}.
  \bibinfo{publisher}{{ACM}}, \bibinfo{pages}{293--302}.
\newblock


\bibitem[\protect\citeauthoryear{Venkatesan and Kumar}{Venkatesan and
  Kumar}{2004}]%
        {venkatesan2004customer}
\bibfield{author}{\bibinfo{person}{Rajkumar Venkatesan} {and}
  \bibinfo{person}{Vita Kumar}.} \bibinfo{year}{2004}\natexlab{}.
\newblock \showarticletitle{A customer lifetime value framework for customer
  selection and resource allocation strategy}.
\newblock \bibinfo{journal}{\emph{Journal of marketing}} \bibinfo{volume}{68},
  \bibinfo{number}{4} (\bibinfo{year}{2004}), \bibinfo{pages}{106--125}.
\newblock


\bibitem[\protect\citeauthoryear{Wang, Wang, Li, Gu, Lu, Zhang, and Gu}{Wang
  et~al\mbox{.}}{2022}]%
        {wang2022cl4ctr}
\bibfield{author}{\bibinfo{person}{Fangye Wang}, \bibinfo{person}{Yingxu Wang},
  \bibinfo{person}{Dongsheng Li}, \bibinfo{person}{Hansu Gu},
  \bibinfo{person}{Tun Lu}, \bibinfo{person}{Peng Zhang}, {and}
  \bibinfo{person}{Ning Gu}.} \bibinfo{year}{2022}\natexlab{}.
\newblock \showarticletitle{CL4CTR: A Contrastive Learning Framework for CTR
  Prediction}.
\newblock \bibinfo{journal}{\emph{arXiv preprint arXiv:2212.00522}}
  (\bibinfo{year}{2022}).
\newblock


\bibitem[\protect\citeauthoryear{Wang, Fu, Fu, and Wang}{Wang
  et~al\mbox{.}}{2017}]%
        {wang2017deep}
\bibfield{author}{\bibinfo{person}{Ruoxi Wang}, \bibinfo{person}{Bin Fu},
  \bibinfo{person}{Gang Fu}, {and} \bibinfo{person}{Mingliang Wang}.}
  \bibinfo{year}{2017}\natexlab{}.
\newblock \showarticletitle{Deep \& cross network for ad click predictions}.
\newblock In \bibinfo{booktitle}{\emph{ADKDD}}. \bibinfo{pages}{1--7}.
\newblock


\bibitem[\protect\citeauthoryear{Wang, Shivanna, Cheng, Jain, Lin, Hong, and
  Chi}{Wang et~al\mbox{.}}{2021}]%
        {wang2021dcn}
\bibfield{author}{\bibinfo{person}{Ruoxi Wang}, \bibinfo{person}{Rakesh
  Shivanna}, \bibinfo{person}{Derek Cheng}, \bibinfo{person}{Sagar Jain},
  \bibinfo{person}{Dong Lin}, \bibinfo{person}{Lichan Hong}, {and}
  \bibinfo{person}{Ed Chi}.} \bibinfo{year}{2021}\natexlab{}.
\newblock \showarticletitle{Dcn v2: Improved deep \& cross network and
  practical lessons for web-scale learning to rank systems}. In
  \bibinfo{booktitle}{\emph{WWW}}. \bibinfo{pages}{1785--1797}.
\newblock


\bibitem[\protect\citeauthoryear{Wang, Liu, and Miao}{Wang
  et~al\mbox{.}}{2019}]%
        {wang2019deep}
\bibfield{author}{\bibinfo{person}{Xiaojing Wang}, \bibinfo{person}{Tianqi
  Liu}, {and} \bibinfo{person}{Jingang Miao}.} \bibinfo{year}{2019}\natexlab{}.
\newblock \showarticletitle{A deep probabilistic model for customer lifetime
  value prediction}.
\newblock \bibinfo{journal}{\emph{arXiv preprint arXiv:1912.07753}}
  (\bibinfo{year}{2019}).
\newblock


\bibitem[\protect\citeauthoryear{Wei, Wang, Li, Nie, Li, Li, and Chua}{Wei
  et~al\mbox{.}}{2021}]%
        {wei2021contrastive}
\bibfield{author}{\bibinfo{person}{Yinwei Wei}, \bibinfo{person}{Xiang Wang},
  \bibinfo{person}{Qi Li}, \bibinfo{person}{Liqiang Nie}, \bibinfo{person}{Yan
  Li}, \bibinfo{person}{Xuanping Li}, {and} \bibinfo{person}{Tat-Seng Chua}.}
  \bibinfo{year}{2021}\natexlab{}.
\newblock \showarticletitle{Contrastive learning for cold-start
  recommendation}. In \bibinfo{booktitle}{\emph{MM}}.
  \bibinfo{pages}{5382--5390}.
\newblock


\bibitem[\protect\citeauthoryear{Win and Bo}{Win and Bo}{2020}]%
        {win2020predicting}
\bibfield{author}{\bibinfo{person}{Than~Than Win} {and}
  \bibinfo{person}{Khin~Sundee Bo}.} \bibinfo{year}{2020}\natexlab{}.
\newblock \showarticletitle{Predicting customer class using customer lifetime
  value with random forest algorithm}. In \bibinfo{booktitle}{\emph{ICAIT}}.
  IEEE, \bibinfo{pages}{236--241}.
\newblock


\bibitem[\protect\citeauthoryear{Wu, Wu, An, Huang, Huang, and Xie}{Wu
  et~al\mbox{.}}{2019}]%
        {wu2019npa}
\bibfield{author}{\bibinfo{person}{Chuhan Wu}, \bibinfo{person}{Fangzhao Wu},
  \bibinfo{person}{Mingxiao An}, \bibinfo{person}{Jianqiang Huang},
  \bibinfo{person}{Yongfeng Huang}, {and} \bibinfo{person}{Xing Xie}.}
  \bibinfo{year}{2019}\natexlab{}.
\newblock \showarticletitle{NPA: neural news recommendation with personalized
  attention}. In \bibinfo{booktitle}{\emph{KDD}}. \bibinfo{pages}{2576--2584}.
\newblock


\bibitem[\protect\citeauthoryear{Xie, Sun, Liu, Wu, Gao, Zhang, Ding, and
  Cui}{Xie et~al\mbox{.}}{2022}]%
        {xie2022contrastive}
\bibfield{author}{\bibinfo{person}{Xu Xie}, \bibinfo{person}{Fei Sun},
  \bibinfo{person}{Zhaoyang Liu}, \bibinfo{person}{Shiwen Wu},
  \bibinfo{person}{Jinyang Gao}, \bibinfo{person}{Jiandong Zhang},
  \bibinfo{person}{Bolin Ding}, {and} \bibinfo{person}{Bin Cui}.}
  \bibinfo{year}{2022}\natexlab{}.
\newblock \showarticletitle{Contrastive learning for sequential
  recommendation}. In \bibinfo{booktitle}{\emph{ICDE}}. IEEE,
  \bibinfo{pages}{1259--1273}.
\newblock


\bibitem[\protect\citeauthoryear{Xing, Bian, Zhao, Xiao, Luo, Yin, Cai, and
  He}{Xing et~al\mbox{.}}{2021}]%
        {xing2021learning}
\bibfield{author}{\bibinfo{person}{Mingzhe Xing}, \bibinfo{person}{Shuqing
  Bian}, \bibinfo{person}{Wayne~Xin Zhao}, \bibinfo{person}{Zhen Xiao},
  \bibinfo{person}{Xinji Luo}, \bibinfo{person}{Cunxiang Yin},
  \bibinfo{person}{Jing Cai}, {and} \bibinfo{person}{Yancheng He}.}
  \bibinfo{year}{2021}\natexlab{}.
\newblock \showarticletitle{Learning Reliable User Representations from
  Volatile and Sparse Data to Accurately Predict Customer Lifetime Value}. In
  \bibinfo{booktitle}{\emph{KDD}}. \bibinfo{pages}{3806--3816}.
\newblock


\bibitem[\protect\citeauthoryear{Yang, Li, and Jobson}{Yang
  et~al\mbox{.}}{2022}]%
        {yang2022personalized}
\bibfield{author}{\bibinfo{person}{Jie Yang}, \bibinfo{person}{Yilin Li}, {and}
  \bibinfo{person}{Deddy Jobson}.} \bibinfo{year}{2022}\natexlab{}.
\newblock \showarticletitle{Personalized Promotion Decision Making Based on
  Direct and Enduring Effect Predictions}.
\newblock \bibinfo{journal}{\emph{arXiv preprint arXiv:2207.14798}}
  (\bibinfo{year}{2022}).
\newblock


\bibitem[\protect\citeauthoryear{Yi, Wang, Ounis, and Macdonald}{Yi
  et~al\mbox{.}}{2022}]%
        {yi2022multi}
\bibfield{author}{\bibinfo{person}{Zixuan Yi}, \bibinfo{person}{Xi Wang},
  \bibinfo{person}{Iadh Ounis}, {and} \bibinfo{person}{Craig Macdonald}.}
  \bibinfo{year}{2022}\natexlab{}.
\newblock \showarticletitle{Multi-modal graph contrastive learning for
  micro-video recommendation}. In \bibinfo{booktitle}{\emph{SIGIR}}.
  \bibinfo{pages}{1807--1811}.
\newblock


\bibitem[\protect\citeauthoryear{Yu, Yin, Xia, Chen, Cui, and Nguyen}{Yu
  et~al\mbox{.}}{2022}]%
        {yu2022graph}
\bibfield{author}{\bibinfo{person}{Junliang Yu}, \bibinfo{person}{Hongzhi Yin},
  \bibinfo{person}{Xin Xia}, \bibinfo{person}{Tong Chen},
  \bibinfo{person}{Lizhen Cui}, {and} \bibinfo{person}{Quoc Viet~Hung Nguyen}.}
  \bibinfo{year}{2022}\natexlab{}.
\newblock \showarticletitle{Are graph augmentations necessary? simple graph
  contrastive learning for recommendation}. In
  \bibinfo{booktitle}{\emph{SIGIR}}. \bibinfo{pages}{1294--1303}.
\newblock


\bibitem[\protect\citeauthoryear{Zhang, Leng, and Liu}{Zhang
  et~al\mbox{.}}{2020}]%
        {zhang2020research}
\bibfield{author}{\bibinfo{person}{Wei Zhang}, \bibinfo{person}{Xuemei Leng},
  {and} \bibinfo{person}{Siyu Liu}.} \bibinfo{year}{2020}\natexlab{}.
\newblock \showarticletitle{Research on mobile impulse purchase intention in
  the perspective of system users during COVID-19}.
\newblock \bibinfo{journal}{\emph{Personal and Ubiquitous Computing}}
  (\bibinfo{year}{2020}), \bibinfo{pages}{1--9}.
\newblock


\bibitem[\protect\citeauthoryear{Zhao, Wu, Tao, Qu, Zhao, Fan, and Zhao}{Zhao
  et~al\mbox{.}}{2023}]%
        {zhao2023percltv}
\bibfield{author}{\bibinfo{person}{Shiwei Zhao}, \bibinfo{person}{Runze Wu},
  \bibinfo{person}{Jianrong Tao}, \bibinfo{person}{Manhu Qu},
  \bibinfo{person}{Minghao Zhao}, \bibinfo{person}{Changjie Fan}, {and}
  \bibinfo{person}{Hongke Zhao}.} \bibinfo{year}{2023}\natexlab{}.
\newblock \showarticletitle{perCLTV: A general system for personalized customer
  lifetime value prediction in online games}.
\newblock \bibinfo{journal}{\emph{TOIS}} \bibinfo{volume}{41},
  \bibinfo{number}{1} (\bibinfo{year}{2023}), \bibinfo{pages}{1--29}.
\newblock


\end{thebibliography}
